\documentclass[ejsv2,noshowframe]{imsart}

\RequirePackage[numbers]{natbib}
\RequirePackage[colorlinks,citecolor=blue,urlcolor=blue]{hyperref}
\RequirePackage{graphicx}
\usepackage[perpage]{footmisc}
\usepackage{amsmath}
\usepackage{amsfonts}
\usepackage{dsfont}
\usepackage{fontenc}
\usepackage{booktabs} 
\usepackage{caption} 
\usepackage{subfigure} 
\usepackage{graphicx}
\usepackage{enumerate}
\usepackage{stmaryrd}
\usepackage{pgfplots}
\usepackage[all]{nowidow}
\usepackage[utf8]{inputenc}
\usepackage{tikz}
\usepackage{tikz-cd}
\usetikzlibrary{babel}
\usetikzlibrary{er,positioning,bayesnet}
\usepackage{multicol}
\usepackage{algorithm2e}
\usepackage{hyperref}
\usepackage[inline]{enumitem} 
\usepackage[linecolor=blue!60!,backgroundcolor=blue!10!,textwidth=25mm,textsize=scriptsize]{todonotes}

\definecolor{blue}{HTML}{1F77B4}
\definecolor{orange}{HTML}{FF7F0E}
\definecolor{green}{HTML}{2CA02C}

\pgfplotsset{compat=1.14}

\newcommand{\Y}{\mathcal{Y}}
\newcommand{\K}{\mathcal{K}}

\newcommand{\R}{\mathbb{R}}
\newcommand{\E}{\mathbb{E}}
\newcommand{\X}{\mathbb{X}}

\newcommand{\B}{\mathcal{B}}

\newcommand{\C}{\mbox{\v{C}}}
\renewcommand{\H}{\mathcal{H}}
\newcommand{\Var}{\operatorname{Var}}
\newcommand{\1}{\mathds{1}}

\renewcommand{\d}{\mathrm{d}}

\arxiv{2303.08456}
\startlocaldefs
\theoremstyle{plain}

\newtheorem{theorem}{Theorem}[section]
\newtheorem{lemma}[theorem]{Lemma}
\newtheorem{corollary}[theorem]{Corollary}
\newtheorem{proposition}[theorem]{Proposition}
\theoremstyle{definition}
\newtheorem{definition}[theorem]{Definition}
\newtheorem*{example}{Example}

\theoremstyle{remark}

\endlocaldefs

\begin{document}
\begin{frontmatter}
\title{Statistical learning on measures: an application to persistence diagrams}
\runtitle{Learning on measures}

\begin{aug}
\author[A]{\fnms{Olympio}~\snm{Hacquard}\ead[label=e1]{hacquard.olympio.4w@kyoto-u.ac.jp}},
\author[B]{\fnms{Gilles}~\snm{Blanchard}\ead[label=e2]{gilles.blanchard@universite-paris-saclay.fr}}
\and
\author[C]{\fnms{Clément}~\snm{Levrard}\ead[label=e3]{clement.levrard@univ-rennes1.fr}}
\address[A]{ASHBi, Kyoto University \printead[presep={,\ }]{e1}}

\address[B]{LMO, Université Paris-Saclay \printead[presep={,\ }]{e2}}

\address[C]{IRMAR, Université de Rennes \printead[presep={,\ }]{e3}}
\runauthor{O. Hacquard et al.}

\end{aug}

\begin{abstract}
We consider a binary supervised learning classification problem where instead of having data in a finite-dimensional Euclidean space, we observe measures on a compact space $\mathcal{X}$. Formally, we observe data $D_N = (\mu_1, Y_1), \ldots, (\mu_N, Y_N)$ where $\mu_i$ is a measure on $\mathcal{X}$ and $Y_i$ is a label in $\{0, 1\}$. Given a set $\mathcal{F}$ of base-classifiers on $\mathcal{X}$, we build corresponding classifiers in the space of measures. We provide upper and lower bounds on the Rademacher complexity of this new class of classifiers that can be expressed simply in terms of corresponding quantities for the class $\mathcal{F}$. If the measures $\mu_i$ are uniform over a finite set, this classification task boils down to a multi-instance learning problem. However, our approach allows more flexibility and diversity in the input data we can deal with. While this general framework has many possible applications, we particularly focus  on classifying data via topological descriptors called persistence diagrams. These objects are discrete measures on $\mathbb{R}^2$, where the coordinates of each point correspond to the range of scales at which a topological feature exists. We will present several classifiers on measures and show how they can heuristically and theoretically enable a good classification performance in various settings in the case of persistence diagrams.
\end{abstract}



\end{frontmatter}

\section{Introduction}

We consider the problem of classifying measures over some metric space. In traditional machine learning, data points are often represented as vectors in Euclidean space, where each dimension corresponds to a feature, making the data suitable for processing by models designed to handle fixed-length numerical inputs. This problem appears as a special instance of standard supervised classification where the data are no longer vectors from a Euclidean space but point clouds or even continuous measures. There are several lines of work looking at this classification task from various perspectives. If the measure is a finite sum of Dirac masses, this problem boils down to multi-instance learning (MIL), where the data are bags of points. This terminology originates in the works from \cite{dietterich1997solving} in the context of drug design. A typical strategy in MIL is to consider a standard classifier over the points from the bag and aggregate the individual labels to classify the entire bag. We refer to the survey from \cite{amores2013multiple} for a comprehensive review of the methods used in multi-instance classification. MIL has since been used as a way to improve machine learning pipelines, for instance by the adjunction of an MIL pooling layer in neural networks \citep{gadermayr2024multiple, deng2024cross} Closer to our work is the paper by \cite{sabato2012multi}, which studies the properties of MIL from a statistical learning perspective. More general are the works on distribution regression initiated by \cite{hein2005hilbertian} for learning on general metric spaces. For instance, \cite{muandet2012learning} tackles the case of classification of distribution and \cite{poczos2013distribution} that of regression. The theory for simple kernel estimators has been developed in \cite{szabo2016learning}. Another recent perspective regarding distribution learning follows the works by \cite{moosmuller2020linear} and \cite{khurana2023supervised}, where the authors consider that each class consists of perturbations of a "mother distribution" and tackle this problem using tools from optimal transport. In the case of input point cloud data, several neural network architectures have been proposed \citep{qi2017pointnet, qi2017pointnet++, zhou2024tnpc}. To conclude our overview of measure-learning methods, we can cite the work from \cite{chazal2021clustering}, where the authors vectorize the measures to cluster them or perform a supervised learning task. The setting we consider is very general in the type of measures we handle, and vectorization-free. We consider simple classifiers based on integrals over the sample measure, and we look at the theoretical performance of such classifiers by relating complexity measures such as Rademacher complexity and covering numbers to their counterparts in the base space. This follows a similar approach as  \cite{sabato2012multi}, while we allow for more general inputs. We can therefore derive generalization error bounds, see \cite{mohri2018foundations} for an introduction to these concepts. We introduce specific classification algorithms which fit into this framework and that discriminate according to the fraction of the mass each measure puts in a well-chosen area.

The theory developed here has many cases of applications, namely, whenever input data are point clouds. We can, for instance, cite LIDAR reconstruction \citealp{de2013unsupervised}, flow cytometry \citep{aghaeepour2013critical}, time series (possibly with an embedding mapping them in some Euclidean space), and text classification using a word embedding method such as \textsc{word2vec}, see \cite{mikolov2013efficient}. Extending the results from MIL, the measures can be weighted depending on the application. We also encompass the case of continuous measures, for example, functional or image classification. 

The main application that motivates the present work is the classification of persistence diagrams. We refer to \cite{edelsbrunner2022computational} for an overview of the construction of this object and of its principal properties. Persistence diagrams are stable topological descriptors of the filtration of a simplicial complex. Mathematically, they are discrete measures on $\mathbb{R}^2$ where both coordinates of each point indicate times at which topology changes occur in the filtration. We can use persistence diagrams to perform various data analysis tasks, and we focus here on supervised classification to discriminate data based on some topological information. Some methods such as landscapes \citep{bubenik2015statistical}, persistence images \citep{adams2017persistence}, or ATOL \citep{chazal2021clustering} immediately get rid of the measure representation and transform the data into a vector. It then becomes possible to plug these vector representations into a standard classifier, we refer to \cite{obayashi2018persistence} for classification using linear classifiers. Some papers use kernel methods, such as \cite{carriere2017sliced} or \cite{le2018persistence}, while some other works make use of neural networks, such as \cite{carriere2020perslay}, and more recently \cite{reinauer2021persformer}. We refer to the survey \cite{hensel2021survey} for an overview of topological machine learning methods. 

In addition to offering a good trade-off between decent predictive performance (comparable to standard persistence diagrams vectorizations and kernel methods) and simplicity, the algorithm developed in
the present work offers explainability guarantees. Indeed, showing that two classes differ on some zones of the persistence diagrams can directly be translated in terms of the range of scales at which relevant topological features exist. The experimental results back up the ideas developed by \cite{bubenik2020persistent} by swiping away a typical paradigm in topological data analysis (TDA), which states that features with a long lifetime are the only ones relevant to describing a shape. Indeed, we demonstrate that the "shape" of the topological noise contains information related to the sampling. This idea is enforced by theoretical guarantees on limiting persistence diagrams as the number of sample points tends to infinity, where we generalize recent results from \cite{owada2021convergence}.

This paper decomposes as follows: in Section \ref{sec:stat_learn} we formalize the problem of learning on a set of measures, give general theoretical guarantees for this problem and propose two simple supervised algorithms. In Section \ref{sec:PD}, we present persistence diagrams that constitute the primary motivation of the present work. We give guarantees on the reconstruction of the proposed algorithm in this specific case, showing that features at every scale can and should be used for classification. Section \ref{sec:expe} contains all the experimental results and the comparison with standard methods, both in TDA and for other applications, showing the versatility of our approach, which we believe is its principal strength, along with its simplicity and explainability. We have made the code publicly available\footnote{ \url{https://github.com/OlympioH/BBA_measures_classification}}. Finally, Section~\ref{sec:proofs} is devoted to the proofs of all the theoretical results contained throughout the paper. 

\section{Statistical learning on measures}

\label{sec:stat_learn}
\subsection{Model}

Let $\mathcal{X}$ be a compact metric space and denote by $\mathcal{M}(\mathcal{X})$ the set of measures of finite mass over $\mathcal{X}$. The model is the following: we observe a sample $D_N = (\mu_i, Y_i)_{i=1}^N$, where $\mu_i \in \mathcal{M}(\mathcal{X})$ and $Y_i$ is a label in $\mathcal{Y} \subset \mathbb{R}$. Although the algorithmic and experimental study is mainly motivated by the case of classification $\mathcal{Y} = \{0, 1\}$, some of the theory developed also encompasses the case of regression where $\mathcal{Y} = [0, 1]$. We aim at building a decision rule $g : \mathcal{M}(\mathcal{X}) \to \mathbb{R}$ that predicts the label $Y^{\prime}$ of a new measure $\mu^{\prime}$. These decision rules are typically built on classes of functions defined on $\mathcal{X}$ itself. There are many practical examples that fall under this framework of learning on a space of measures: functional regression \citep{ferraty2006nonparametric} and image classification are standard examples that have given birth to a very wide variety of problems. Classifying bags of points has been studied under the MIL terminology, we refer to \cite{amores2013multiple} for a complete survey, and cover many useful applications, from which we can cite image classification based on a finite number of descriptors as done in \cite{wu2015deep}, flow cytometry (see Section \ref{sec:expe}), or text classification where each word is represented by a point in a high-dimensional space. Closer to us is the work by \cite{chazal2021clustering} where they represent the measures in a Euclidean space and use these vectorizations to cluster the data. Even though the applications are very similar, we believe that our work is quite different in essence since our algorithms are formulated in a supervised setting, and we do not represent the measures in a Euclidean space, preferring to develop a theory directly for an input space of measures, as we will see in the following section.

\subsection{Theoretical complexity bounds}
\label{sec:theory}
In this section, we adapt standard results in statistical learning theory by relating quantities such as Rademacher complexity and covering numbers for functional classes over $\mathcal{X}$ to their counterparts in the space $\mathcal{M}(\mathcal{X})$. In what follows, $\mathcal{R}_N(\cdot)$ (resp. $G_N(\cdot)$) denotes the empirical Rademacher (resp. Gaussian) complexity of a function class conditionally on a sample $(Z_1, \ldots, Z_N)$ which we recall is defined by
\[ \mathcal{R}_N (\mathcal{F}) = \frac{1}{N} \mathbb{E}_\sigma \left[ \underset{f \in \mathcal{F}} \sup \left| \sum_{i=1}^N \sigma_i f(Z_i) \right| \right], 
\]
where $(\sigma_1, \ldots, \sigma_N)$ are independent Rademacher random variables. The Gaussian complexity obeys the same definition where the $\sigma_i$ are independent standard normal variables. The Rademacher complexity is a usual quantity in statistical learning that measures the richness of a set of functions. Loosely speaking, it quantifies how much the class $\mathcal{F}$ correlates with a vector of noise $(\sigma_1, \ldots, \sigma_N)$. This quantity naturally appears when controlling the performance of a family of classifiers; a large Rademacher complexity being detrimental to a good generalization. We refer to Chapter 26 of \cite{shalev2014understanding} for more details. It is common to upper bound this
quantity by computing the covering number of the function class. We denote by $\mathcal{N}(\mathcal{F}, d, \varepsilon)$ (resp. $\mathcal{M}(\mathcal{F}, d, \varepsilon)$) the $\varepsilon$-covering (resp. packing) number of the set $\mathcal{F}$ endowed with metric $d$. Finally, we denote by $\mathrm{VC}(\mathcal{F})$ the Vapnik-Chervonenkis dimension of a set of functions (or its pseudo-dimension in the case of real hypotheses classes) and by $\mathrm{VC}(\mathcal{F}, \gamma)$ its $\gamma$-fat shattering dimension. We refer to Chapter 6 of \cite{shalev2014understanding} for the definition of these concepts that measure the capacity of a function class, and are also used to upper bound the validation error of a classification model. We break down our analysis in two cases: the first one assumes that we have discrete finite measures and that we apply the 0-1 loss while the second assumes generic measures inputs and requires the loss function to be Lipschitz. 

\subsubsection{Discrete measures, 0-1 loss}
We denote by $\mathcal{M}_m (\mathcal{X})$ the set of measures that write as a finite sum of at most $m$ Dirac masses on $\mathcal{X}$, i.e. $\mu_i = \sum_{j=1}^{n_i} \delta_{x_j^i}$ with $n_i \leq m$ for all $i$. We consider a family $\mathcal{F}$ of classifiers from $\mathcal{X}$ to $\{0, 1\}$. For a given $f \in \mathcal{F}$, we have a set of predictions for each individual point:

\[f(\mu_i) = [f(x_1^i), f(x_2^i), \ldots, f(x_{n_i}^i)] \in \{0, 1\}^{n_i}.
\]

Denoting by $\{0, 1 \}^\star$ the set of finite sequences of $0$'s and $1$'s, we finally apply some function $\psi : \{0, 1 \}^\star \to \{0, 1 \}$ called a bag-function or an aggregation function in order to output a prediction label for each measure. Described as such, this scenario is formulated exactly as a Multi-Instance Learning (MIL) problem, and theoretical guarantees in this case have been established by \cite{sabato2012multi}. In particular, the authors provide upper-bounds on the VC-dimension, covering numbers, and Rademacher complexity of such \textit{bag of points} classifiers. In Proposition \ref{prop:VC}, we extend their results, in particular their Theorem 6 to the case where $\psi$ is no longer a fixed function but is itself learned from a VC-class $\mathcal{G}$. Assume the bag-function $\psi$ to be permutation invariant, i.e. $ \psi(y_1, \ldots, y_n) = \psi(y_{\sigma(1)}, \ldots, y_{\sigma(n)})$ for every $y_i \in \{0, 1\}, n \in \mathbb{N}$, and $\sigma \in \mathfrak{S}_n$. Then there exist two functions $g$ and $\bar{\psi}$ such that $\psi$ decomposes as follows:

\[
\begin{tikzcd}[column sep=huge,row sep=huge]
\{0, 1\}^{\star} \arrow[r,"\psi = \bar{\psi} \circ g"] \arrow[dr,swap,"g"] &
  \{0, 1 \} \\
& \mathbb{R}^2 \arrow[u,shift left=.75ex,"\bar{\psi}"]
\end{tikzcd}
\]

The function $g$ is defined as $g(y_1, \ldots, y_n) = (\sum_{i=1}^n y_i /n, n)$, i.e. it maps a sequence of zeros and ones to the proportion of ones and the total number of elements in the sequence. We denote by $\mathcal{H}$ the set of binary classifiers from $\mathcal{M}_m(\mathcal{X})$ defined as $h: \mu = \sum_{i=1}^{n} \delta_{x_i} \mapsto \psi(f(x_1), \ldots, f(x_{n}))$, where $\psi \in \mathcal{G}$ and $f \in \mathcal{F}$.

\begin{proposition}
\label{prop:VC}
Assume all the input measures belong to $\mathcal{M}_m (\mathcal{X})$. Assume $\psi$ is taken from a class $\mathcal{G}$ of permutation invariant functions and that the corresponding $\bar{\psi}$ is taken from a class $\bar{\mathcal{G}}$ of VC-dimension $d^\prime$. We further assume that the class $\mathcal{F}$ has a finite VC-dimension $d$. Then, $\mathcal{H}$ is a VC-class of dimension $d_2$ verifying:

\[d_2 \leq \max(16, 2(d+d^\prime) \log_2 (2 e m)).
\]

\end{proposition}

We defer the proof to Section \ref{proof:VC}. We observe that in the case where $d^\prime = 0$, this is equivalent to the Theorem 6 of \cite{sabato2012multi}. This bound on the VC dimension of the composition of a hypothesis class $\mathcal{F}$ with a class of bag-functions can be used to upper-bound the classification accuracy of predictors over the set of measures $\mathcal{M}_m(\mathcal{X})$. We now propose to extend these results to the case of general measures with finite mass and therefore extend the MIL framework.

\subsubsection{Generic measures, Lipschitz loss}
In this section, using $\mathcal{Y} = \{-1,1\}$ we build classifiers of the form $\text{sgn}(g (\mu))$ for $g$ in some function class $\mathcal{G}$ over $\mathcal{M}(\mathcal{X})$. Consider a $\kappa$-Lipschitz loss function $\mathcal{L}$. By the contraction principle for Rademacher complexities, it holds $\mathcal{R}_N (\mathcal{L}\circ \mathcal{G}) \leq \kappa \mathcal{R}_N (\mathcal{G})$. We therefore focus on the control of the Rademacher complexity of the class of real-valued predictors. We first extend Lemma 12 from \cite{sabato2012multi} to our setting. In what follows, we consider a class of functions $\mathcal{F}$ from $\mathcal{X}$ to 
$[0, 1]$, and the associated class of functions $\tilde{\mathcal{F}}$ defined on $\mathcal{M}(\mathcal{X})$ by 
\[\tilde{f}[\mu] = \mathbb{E}_{X \sim \mu} [f(X)] = \int_{\mathcal{X}} f(x) \mathrm{d}\mu(x) \text{ for } f \in \mathcal{F}.
\] 
The following lemma gives a relationship between the covering numbers of $\mathcal{F}$ and $\tilde{\mathcal{F}}$. We denote by $L_p^N$ the in-sample $p$-norm, defined for two functions $f_1$ and $f_2$ in $\mathcal{F}$, and a sample of $N$ measures $(\mu_1, \ldots, \mu_N)$ as:

\[\|\tilde{f_1}-\tilde{f_2}\|_{L_p^N} = \left( \frac{1}{N} \sum_{i=1}^N ( \tilde{f_1}[\mu_i] - \tilde{f_2}[\mu_i] )^p \right)^{1/p}.
\]

Given a sample $(\mu_1, \ldots, \mu_N) \in \mathcal{M}(\mathcal{X})^N$, we denote by $\bar{M}_p = \left( \frac{1}{N} \sum_{i=1}^N M_i^{p} \right)^{1/p}$ where $M_i = \mu_i(\mathcal{X})$ is the total mass of the measure $\mu_i$.

\begin{lemma}
\label{lemma:covering}
Let $(\mu_1, \ldots, \mu_N) \in \mathcal{M}(\mathcal{X})^N$ and let $p \in [1, + \infty].$ There exists a probability measure $\bar{\mu}$ such that

\[\mathcal{N}(\tilde{\mathcal{F}}, L_p^N, \varepsilon) \leq \mathcal{N} \left(\mathcal{F}, L_p (\bar{\mu}), \frac{\varepsilon}{\bar{M}_p} \right).
\] 

\end{lemma}

We defer the proof to Section \ref{sec:proofs}. This can be used to upper-bound the Rademacher complexity of the function class $\tilde{\mathcal{F}}$ as shown in the following theorem.

\begin{theorem}
\label{rad_sup}
There exists an absolute constant $K$ such that
\[\mathcal{R}_N(\tilde{\mathcal{F}}) \leq \frac{K \bar{M}_2 \sqrt{\mathrm{VC}(\mathcal{F})}}{\sqrt{N}}.
\]
\end{theorem}

For $M >0$, we define the set $\mathcal{C}_M^N = \{ (\mu_1, \ldots, \mu_N) \in (\mathcal{M}(\mathcal{X}))^N | \frac{1}{N} \sum_{i=1}^N \mu_i^2 (\mathcal{X}) = M^2 \}$. In addition to the previous theorem, we provide a lower bound of the same order for the Rademacher complexity.

\begin{theorem}
\label{rad_inf}
There exists an absolute constant $K^{\prime}$ such that

\[ \frac{K^{\prime} \bar{M}_2}{ \sqrt{N} \ln(N)}  \sqrt{\mathrm{VC}(\mathcal{F})}\leq \underset{(\mu_1, \ldots, \mu_N) \in \mathcal{C}_{\bar{M}_2}^N}{\sup} \mathcal{R}_N \left( \tilde{\mathcal{F}}\Big|\mu_1, \ldots, \mu_N \right).
\]

\end{theorem}

The bounds from Theorems \ref{rad_sup} and \ref{rad_inf} match and are both of order $1/\sqrt{N}$, up to logarithmic factors. They also both depend on the VC-dimension of the base-class $\mathcal{F}$ and no longer of $\tilde{\mathcal{F}}$, making it much easier to compute, as we can see in the example below.

\begin{example}	
Assume $\mathcal{X}$ is a bounded subspace of $\mathbb{R}^d$ endowed with a Euclidean metric and let $\mathcal{F} = \{ \mathds{1}_{\mathcal{B}(x, r)} | x \in \mathcal{X}, r>0 \}$. It is a standard fact (see \cite{mohri2018foundations} for instance) that the VC-dimension of Euclidean balls is $d+1$. We therefore have by Theorem \ref{rad_sup} that there exist constants $K$ and $K^\prime$ such that:

\[ \frac{K^{\prime} \bar{M}_2 \sqrt{d+1}}{ \sqrt{N} \ln(N)} \leq \underset{(\mu_1, \ldots, \mu_N) \in \mathcal{C}_{\bar{M}_2}^N}{\sup} \mathcal{R}_N \left( \tilde{\mathcal{F}}\Big|\mu_1, \ldots, \mu_N \right) \leq \frac{K \bar{M}_2 \sqrt{d+1}}{\sqrt{N}}.
\]

\end{example}

In practice, the class $\tilde{\mathcal{F}}$ is used to construct a binary classifier through composition with an aggregation function $\psi$, whose sign gives a prediction in $\{-1,1\}$. If the function $\psi$ is fixed as it is the case in \cite{sabato2012multi} and is further assumed to be $L$-Lipschitz, the Rademacher complexity of the final set of classifiers is simply multiplied by $L$. We want to generalize this to the case where the function $\psi$ is also learned. Assume $\psi : \mathbb{R} \to \mathbb{R}$ is taken from a class of functions $\mathcal{G}$. Denote by $\mathcal{H}$ the class of functions $ h : \mu \mapsto \psi\big(\int_\mathcal{X} f(x) \mathrm{d} \mu(x)\big)$ where $f \in \mathcal{F}, \psi \in \mathcal{G}$. The following proposition gives a bound on the Gaussian complexity of the function class $\mathcal{H}$.

\begin{proposition}	
\label{prop:comp}
Assume that the class $\mathcal{G}$ consists of $L$-Lipschitz functions. Assume the null function $x \mapsto 0$ belongs to $\mathcal{F}$. Then there exist constants $C_1$ and $C_2$ such that for any sample of measures $\bar{\mu} = (\mu_1, \ldots, \mu_N)$,

\[G_N(\mathcal{H}) \leq \frac{C_1 \bar{M}_2 L \sqrt{\mathrm{VC}(\mathcal{F})} \sqrt{\log(N)}}{\sqrt{N}} + \frac{C_2 L \bar{M}_2 \mathbf{R} (\mathcal{G})}{\sqrt{N}} + \frac{L}{\sqrt{N}} \underset{\psi \in \mathcal{G}}{\sup} |\psi(0)|,
\]
where 
\[\mathbf{R}(\mathcal{G}) = \sup _{\mathbf{x}, \mathbf{x}^{\prime} \in \mathbb{R}, \mathbf{x} \neq \mathbf{x}^{\prime}} \mathbb{E}_{\gamma} \sup _{\psi \in \mathcal{\mathcal{H}}} \frac{(\psi(\mathbf{x})-\psi(\mathbf{x^\prime))} \gamma}{|\mathbf{x}-\mathbf{x}^{\prime}|},
\] 
and where $\gamma \sim \mathcal{N} (0, 1)$.

\end{proposition}

We refer to Section \ref{proof:comp} for the proof. This proposition shows that up to logarithmic factors, the Gaussian complexity of the family of classifiers decreases at an overall rate of $1/\sqrt{N}$. The quantity $\mathbf{R}(\mathcal{G})$ appears as a supremum of Gaussian averages. We refer to Theorem 5 of \cite{maurer2016chain} for a few properties of this quantity. Most notably, if the class $\mathcal{G}$ is finite, and consists of $L$-Lipschitz functions, $\mathbf{R}(\mathcal{G}) \leq L \sqrt{2 \ln | \mathcal{G} |}$. In addition, in some simple cases, it is possible to provide a better estimate of $\mathbf{R} (\mathcal{G})$, even when $\mathcal{G}$ is infinite, as we can see in the following example:

\begin{example}
\label{ex:thresh}
In practice, we often choose $\psi$ of the form $\psi : x \mapsto x-s$ where $s \in [-S, S]$ is learned, as we will see in Section \ref{sec:algo}. In this case, we directly have that $\mathbf{R}(\mathcal{G}) = \mathbb{E} [| \gamma |] = 1.$ Therefore, keeping the same notation as above, in this scenario there exist universal constants $C_1$ and $C_2$ such that

\[G_n(\mathcal{H}) \leq  \frac{1}{\sqrt{N}} \left[ C_1  \bar{M}_2 \sqrt{\mathrm{VC}(\mathcal{F})} \sqrt{\log(N)}+S+C_2 \bar{M}_2  \right].
\]

\end{example}

\subsection{Algorithms, application to rectangle-based classification}
\label{sec:algo}

Let us consider a class $\mathcal{A}$ of Borel sets of $\mathcal{X}$. For instance, $\mathcal{A}$ can be thought of as the set of balls or axis-aligned hyperrectangles for a given metric. We then consider the class of corresponding indicator functions $\mathcal{F} = \{ \mathds{1}_{A}, A \in \mathcal{A} \}$. The data are therefore classified given some threshold $s \in \mathbb{R}$ and a sign $\varepsilon \in \{-1, +1\}$, by the decision rule $\mu \mapsto \mathds{1} \big\{\varepsilon \mu(A)
- s \geq 0 \big\}$.

If $\mathcal{A}$ is a set of balls, the optimization problem boils down to finding the best center in $\mathcal{X}$ and the best radius in $\mathbb{R}_{+}$. We present two algorithms and associate each of them with the theory developed in the previous subsection. 

\subsubsection*{Algorithm 1: exhaustive search}
The first method consists in performing an exhaustive search in a discretized grid of parameters for a threshold $s \geq 0$ and
for the set $\mathcal{A}$, and select
those that minimize the empirical classification error:
\begin{equation}
\label{eq:empir_risk}
  (A^+,t^+) = \mathop{\rm Arg\,Min}_{A,t} \mathcal{L}_{+}(A,t),
\end{equation}
where
\[\mathcal{L}_{+}(A,t) = \sum_{i=1}^N \mathds{1}\left\{\int_{A} \mathrm{d} \mu_i -t > 0 \right\} \mathds{1} \{ Y_i = 0 \} + \mathds{1}\left\{\int_{A} \mathrm{d} \mu_i -t \leq 0 \right\} \mathds{1} \{ Y_i = 1 \}.
\]
We similarly minimize the empirical classification error for reversed labels: $(A^-,t^-) = \mathop{\rm Arg\,Min}_{A,t} \mathcal{L}_{-}(A,t)$, for
\[\mathcal{L}_{-}(A,t) = \sum_{i=1}^N \mathds{1}\left\{\int_{A} \mathrm{d} \mu_i -t \leq 0 \right\} \mathds{1} \{ Y_i = 0 \} + \mathds{1}\left\{\int_{A} \mathrm{d} \mu_i -t > 0 \right\} \mathds{1} \{ Y_i = 1 \}.
\]
If $ \mathcal{L}_{+}(A^+,t^+) \leq \mathcal{L}_{-}(A^-,s^-)$ we set $\varepsilon = 1$ and pick $(A^+,t^+)$, otherwise we set $\varepsilon = -1$, along with the corresponding set of parameters.

If all the measures $\mu_i$ write as a finite sum of Dirac measures, this step is very similar to MIL, since each of the $N_i$ points in the bag $\mu_i$ will be assigned a label according to whether it belongs to the set $A$ or not. The additional component is that we consider multiple aggregation functions of the form 

\begin{align*}
  \psi_s \colon \{0, 1 \}^{N_i} &\to \{0, 1\}\\
  x &\mapsto \mathds{1} \left\{ \sum_{j=1}^{N_i} x_j \geq s \right\}.
\end{align*}

Here, we allow the threshold $s$ to be learned, which extends the theory developed in Chapter 3 of \cite{sabato2012multi} about binary MIL where the aggregation function must be fixed. We therefore fit exactly within the framework of Proposition \ref{prop:VC} provided that the set of raw classifiers $\mathcal{F} = \{ \mathds{1}_{A} |A \in \mathcal{A} \}$ is a VC-class, which is for instance the case if $\mathcal{A}$ is a set of Euclidean balls or axis-aligned hyper-rectangles. Note that this algorithm allows for any sample of measures with finite mass as input.

\subsubsection*{Algorithm 2: smoothed version}
Performing an exhaustive search has a computational cost that grows exponentially with the dimension of the space in which the data lie. We propose to optimize a smoothed version of the empirical error. In the case of balls, for a center $C \in \mathcal{X}$, a radius $r>0$, a threshold $s$ and a scale $\sigma$, we consider the predictor given by the sign of $f_{C, r, s, \sigma}$, defined as

\[f_{C, r, s, \sigma} (\mu) = \int_\mathcal{X} \exp \left(-\frac{d(\mathcal{B}(C, r), x)}{\sigma} \right) \mathrm{d} \mu(x) - s.
\]
We minimize the cross-entropy loss between a smooth version of this predictor and the target vector, for a sample $D_N = (\mu_i, Y_i)_{i=1}^N$:

\[ \mathcal{L}_{D_N}(C, r, s, \sigma) = -\sum_{k=1}^N Y_k \log(P(f_{C, r, s, \sigma} (\mu_k)) + (1-Y_k) \log(1-P(f_{C, r, s, \sigma} (\mu_k)),
\]
where $P$ is the sigmoid function: $x \mapsto \frac{1}{1+e^{-x}}$. This optimization must be performed for switched labels as well.

In practice, we perform a stochastic gradient descent of this loss function. Since this objective typically has many critical points, we perform multiple runs with different initialization parameters.

The predictor $P \circ f_{C, r, s, \sigma}$ is a smooth predictor that has output in $\mathcal{Y}=[0, 1]$. This algorithm can also be interpreted using the MIL lens if the $\mu_i$'s are discrete sums of Dirac measures. Indeed, the class of functions we consider is 
\[
\mathcal{F} = \left\{ x \mapsto \exp \left(-\frac{d(\mathcal{B}(C, r), x)}{\sigma} \right) \Big| C \in \mathcal{X}, (r, \sigma) \in (\mathbb{R}_{+} )^2 \right\},
\] 
so that each point in the bag $\mu_i$ is mapped to a real number which corresponds to the framework of Section 6.2 of \cite{sabato2012multi}. The class $\mathcal{F}$ is a smoothed version of ball indicators and has the same VC dimension: $\mathrm{VC}(\mathcal{F}) = d+1$. Using Proposition \ref{prop:comp} with the class $\mathcal{G} = \{ x \mapsto P(x-s) \}$, we can therefore write the corresponding generalization bound, using that the cross-entropy loss is $1$-Lipschitz. According to Theorem 26.5 of \cite{shalev2014understanding}, we have that with probability at least $1- \delta$, for all $\theta = (C, r, \sigma, s) \in \mathcal{X} \times \mathbb{R}_{+}^2 \times [-S, S]$, 
\[\mathbb{E}_{D_N} [\mathcal{L}_{D_N}(\theta)] - \mathcal{L}_{D_N} (\theta) \leq \sqrt{\frac{\pi}{2N}} \left[ C_1 \sqrt{(d+1) \log(N)}\bar{M}_2+S+C_2 \bar{M}_2 + C_3 \sqrt{\log(4/\delta)}  \right],
\]
with universal constants $C_1$, $C_2$ and $C_3$.

\subsubsection*{Aggregation by boosting}

The two methods presented above select a single Borel set to discriminate between the two classes. This approach suffices when the two classes always differ in the same zone of $\mathcal{X}$ but has obviously limited expressivity capabilities. 
We therefore propose in practice to combine this base ``weak learner'' with a boosting approach. 
We have implemented the \textsc{adaboost} method \citep{freund1996experiments}, which classically calls the base method
iteratively, giving more weight to misclassified data. In addition to greatly improving the predictive performance as opposed to selecting a single convex set, performing boosting is of qualitative interest since it shows which zones of the measures are relevant for classification. This feature is of particular interest in applications where these areas convey a qualitative information, such as persistence diagrams or flow cytometry.

\section{A leading case study: classifying persistence diagrams}
The primary example of measures that motivates the present work are persistence diagrams and their smoothed and weighted variants.
\label{sec:PD}
\subsection{An introduction to persistence diagrams}
Persistence diagrams are measures on $\mathbb{R}^2$ that summarize the topological properties of input data and constitute one of the main objects in Topological Data Analysis (TDA). We refer to the textbooks by \cite{edelsbrunner2022computational} and \cite{boissonnat2018geometric} for complete and exhaustive treatment, and hereby recall the principal notions. We focus on the simplest discrete framework of \textit{simplicial complexes}:

\begin{definition}
A \textit{(finite) abstract simplicial complex} $\mathcal{K}$, or \textit{simplicial complex}, is a finite collection of finite sets that is closed under taking subsets. An element $\sigma\in\mathcal{K}$ is called a \textit{simplex}, and subsets of $\sigma$ are called \textit{faces} of $\sigma$. The dimension of a simplicial complex is the maximal dimension of one of its simplices.
\end{definition}

It can be shown, see Chapter III of \cite{edelsbrunner2022computational}, that every simplicial complex of dimension $d$ has a geometric realization in $\mathbb{R}^{2d+1}$ obtained by mapping the vertices to $\mathbb{R}^{2d+1}$. A \textit{(geometric) $k$-dimensional simplex} is the convex hull of $k+1$ affinely independent points. A $0$ (resp. $1$, $2$, $3$)-dimensional simplex is called a vertex (resp. an edge, a triangle, a tetrahedron). Given a simplicial complex $\K$ of dimension $d$, its Betti numbers $\beta_0, \ldots, \beta_{d-1}$ are topological invariants that correspond to the number of $k$-holes (i.e. $\beta_0$ is the number of connected components, $\beta_1$ of cycles, and so on). We refer to \cite{edelsbrunner2022computational} for a rigorous introduction. The sequence of all Betti numbers helps characterizing the topology of a simplicial complex. One of the principal examples of geometric simplicial complexes built over a point cloud is the \v{C}ech complex:

\begin{definition}
\label{def:Cech}
Let $\X\subseteq \R^d$ be finite. The \textit{\v{C}ech complex at scale~$r\geq 0$} is the simplicial complex $\C(\X, r)$ defined as follows: for $(x_0, \ldots, x_{k}) \in \X^{k+1}$, the simplex~$[x_0, \ldots, x_k]$ is in $\C(\X, r)$ if the intersection of closed balls $\cap_{l=0}^{k} \bar{B}(x_l, t)$ is non-empty.
\end{definition}

The key to persistence theory is to consider simplicial complexes with a multi-scale approach and consider a sequence of nested complexes rather than a single complex. To that extent, we can define a \textit{filtration} of a simplicial complex as follows:

\begin{definition}
\label{def:filtration}
Consider a finite simplicial complex $\K$ and a non-decreasing function $f: \K \to \R$, in the sense that $f(\sigma) \leq f(\tau)$ whenever $\sigma$ is a face of $\tau$. We have that for every $a \in \mathbb{R}$, the sublevel set $\K(a) = f^{-1}(-\infty, a]$ is a simplicial subcomplex of $\K$. Considering all possible values of $f$ leaves us with a nested family of subcomplexes
\[\emptyset = \K_0 \subset \K_1 \subset \ldots \subset \K_n = \K,
\]
called a \textit{filtration}, where $a_0 = -\infty <a_1 <a_2 < \ldots <a_n$ are the values taken by $f$ on the simplices of $\K$.
\end{definition}

For instance, considering all possible scales of the \v{C}ech complex of a point set $\X$ naturally defines a filtration over the complete simplicial complex $\K=2^{\X}$. The \v{C}ech filtration of $\X$, denoted by $\C(\X)$ is equivalent to centering balls around each point of $\X$ and have the balls' radii grow from $0$ to $\infty$. For general filtrations, as the scale parameter grows we are interested in tracking the evolution of the Betti numbers of the simplicial complexes. If a $k$-dimensional hole starts to exist at some time $b$ and disappears at time $d$ in the filtration, we add the point $(b, d)$ in the $k$-dimensional persistence diagram $D_k$ of the filtration. A persistence diagram therefore appears as a multi-set of points supported in the half-plane $H$ defined by $H=\{(x, y) \in \mathbb{R}^2 |  \hspace{0.5em} x \leq y \leq +\infty \}.$ We can equivalently look at persistence diagrams as discrete measures on $H$: $\xi_k = \sum\limits_{(b,d) \in D_k} \delta_{(b,d)}$ to conform to the theory and algorithms developed in Section \ref{sec:theory}.

\begin{figure}[t]
\begin{center}
\subfigure[$n=100$]{\includegraphics[scale=0.3]{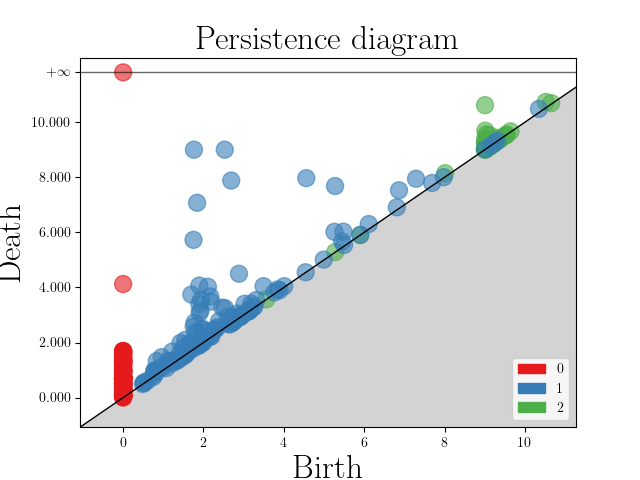}}
\subfigure[$n=200$]{\includegraphics[scale=0.3]{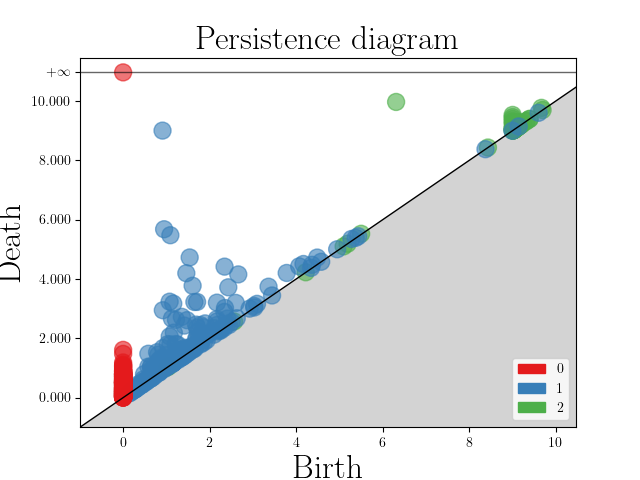}} \\
\subfigure[$n=500$]{\includegraphics[scale=0.3]{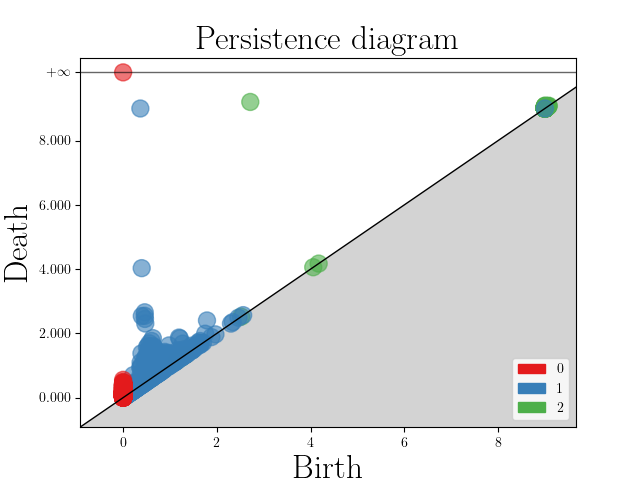}}
\subfigure[$n=1000$]{\includegraphics[scale=0.3]{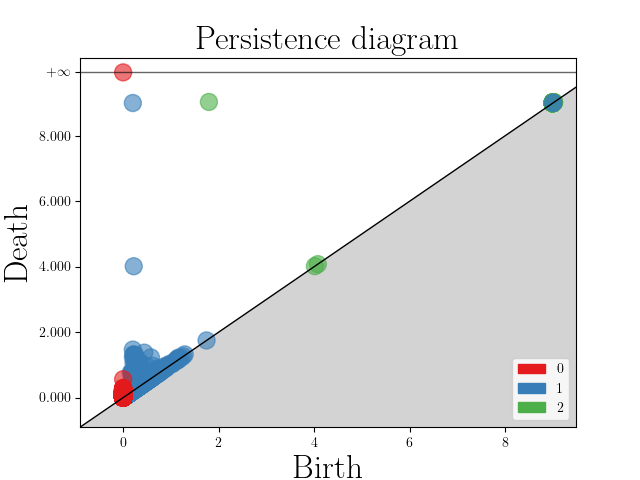}}
\end{center}
\caption{0, 1 and 2-persistence diagrams for $n$ points uniformly sampled on a torus.}
\label{dgm_ex_torus}
\end{figure}

We illustrate the construction of $0$, $1$ and $2$-persistence diagrams of a \v{C}ech filtration in Figure \ref{dgm_ex_torus} where we sample $n$ points uniformly on a torus, according to an algorithm provided by \cite{diaconis2013sampling}. When the number of points is very low ($n=100$), the true homology of the manifold (one feature of dimension 2 and two features of dimension 1) does not show in the diagrams and we only observe topological components due to the sampling. For $n=500$, we can read the homology of the torus in the persistence diagram along with many points close to the diagonal. As $n$ grows, this "topological noise" concentrates around the origin and the true homological features become well separated from the noise. If we sample a point cloud from a manifold, large-persistence features correspond to proper homological features of the manifold, see Theorem~\ref{theo:BD}. Following this approach, works such as \cite{adams2017persistence} on persistence images suggest weighting the persistence diagram using an increasing function of the persistence. In addition, they propose to convolve the discrete measure with a Gaussian function. This falls under the framework of the previous section, and it becomes relevant to consider diagrams as generic measures. However, this signal-noise dichotomy is very restrictive, and there is some evidence that points lying close to the diagonal also carry relevant information such as curvature as demonstrated in \cite{bubenik2020persistent} or dimension. We give further evidence of that claim in the following section, where we show that asymptotically, we can extract information on the sampling density around the origin of the limiting persistence diagram. We also provide numerical illustrations and quantitative evidence that low-persistence features are relevant for classification purposes. 

Finally, note that the success of persistence diagrams in topological data analysis has been motivated by the possibility to compare diagrams (with possibly different number of points) using distances inspired by optimal transport, the most popular being the \textit{bottleneck distance}, which benefits from some stability properties \citep{cohen2005stability}:
\begin{definition}
Let $\Delta = (x, x) \subset H$ be the diagonal of $\R^2$.

The \textit{bottleneck distance} $d_B$ between two persistence diagrams $D$ and $D^\prime$ is defined by:

\[ d_B (D, D^\prime) = \underset{ \eta : D \cup \Delta \to D^\prime \cup \Delta} { \inf} \quad \underset{x \in D \cup \Delta} {\sup} \| x- \eta (x) \|_\infty,
\] 
where the infimum is taken over all bijections $\eta$ from $D \cup \Delta$ to $D^\prime \cup \Delta$. 
\end{definition}

\subsection{Structural properties of persistence diagrams}
Throughout this section and the following, we consider classifiers constructed by finding the best axis-aligned rectangle. The easiest information to capture on the persistence diagram of a \v{C}ech complex is the global one corresponding to the homology of the manifold supporting the data. If we consider samplings on metric spaces having different persistence diagrams for a given filtration, the following theorem yields the existence of a rectangle classifier that discriminates between the two supporting spaces with high probability. Before stating the theorem, we recall the definition of an $(a,b)$-standard measure:

\begin{definition}
Let $\mathbb{X}$ be a compact metric space and let $a, b >0$. We say that a probability measure $\mu$ on $\mathbb{X}$ satisfies the \textit{$(a,b)$-standard assumption} if

\[\forall x \in \mathbb{X}, \forall r>0, \mu (\mathcal{B}(x, r)) \geq \min (1, ar^b).
\]
\end{definition}

In the following theorem, $\mathrm{dgm}$ denotes a persistence diagram (of any homological dimension) of a filtration.

\begin{theorem}
\label{theo:BD}
Let $\mathbb{M}_1$ and $\mathbb{M}_2$ be two compact metric spaces. Assume that we observe an i.i.d. sample $\hat{X}_n = (X_i)_{i=1}^n$ drawn from an $(a_1, b_1)$-standard measure on $\mathbb{M}_1$ or an $(a_2, b_2)$-standard measure on $\mathbb{M}_2$. Denote by $K_i = \min\limits_{(p,q) \in \mathrm{dgm} \left(\C(\mathbb{M}_i)\right)} |q-p|$ for $i=1, 2$. Assume that there exists $K_3>0$ such that

\[d_B (\mathrm{dgm}(\C(\mathbb{M}_1)), \mathrm{dgm}(\C(\mathbb{M}_2))) \geq K_3.
\]
Denote by $K=\min (K_1, K_2, K_3), a= \min(a_1, a_2)$ and $b= \min(b_1, b_2)$. For all $\delta >0$, if the number $n$ of sample points verifies

\[n \geq \frac{2^b}{aK^b} \log \left(\frac{4^b}{aK^b \delta} \right),
\]
there exists a collection of rectangles for which the classification error is smaller than $\delta$.

\end{theorem}

We defer the proof to Section \ref{proof:BD}. We refer to \cite{chazal2014persistence} for the construction of the \v{C}ech filtration of possibly infinite metric spaces (and not simplicial complexes as previously), which ensures that $\mathrm{dgm}(\C(\mathbb{M}_1))$ and $\mathrm{dgm}(\C(\mathbb{M}_2))$ are well defined.

In practice, a lot of information is contained in the points lying close to the diagonal and classifying persistence diagrams enables to deal with a far broader class of problems than simply classifying between manifolds with different homology groups, as we will see in the following sections. Studying geometric quantities of the \v{C}ech complex of a random point cloud $\X_n = (X_i)_{i=1}^n$ on $\mathbb{R}^d$ is a deeply studied problem and we refer to \cite{bobrowski2011distance} for some preliminary results regarding the critical points of the distance function to a point cloud. Some results have been adapted in \cite{bobrowski2015topology} to point clouds sampled on manifolds and convergence results in the Wasserstein sense have been established in \cite{arnal2024wasserstein}. Assume that we consider the \v{C}ech complex at a scale $r_n$ that decreases with $n$ and such that $r_n \to 0$. The speed at which $r_n$ tends to 0 as $n \to \infty$ is paramount and dictates the type of results we can expect. In what follows, we focus on the \textit{sparse regime}, i.e. $nr_n^d \to 0$ as $n \to \infty$. In this regime, asymptotic properties of the persistence diagram of the \v{C}ech filtration have been studied in \cite{owada2021convergence}. When considering persistent quantities, we consider the \v{C}ech complex at all possible ranges and we renormalize the sample points themselves by the sequence $r_n$. We generalize the results of the above-mentioned citation in the following theorem where our contribution is two-fold: the data are now allowed to be sampled from a manifold, and we provide a rate of convergence of the persistence diagram towards its limiting measure. Before stating the theorem, we define the function $h_r$ by
\[h_r\left(x_1, \ldots, x_{k+2}\right)=\mathds{1}\left\{\beta_k\left(\mbox{\v{C}}\left(\left\{x_1, \ldots, x_{k+2}\right\}, r\right)\right)=1\right\},
\] 
and for $0 \leq s \leq t \leq u \leq v \leq \infty$,
\[H_{s, t, u, v}(\mathbf{x})=h_t(\mathbf{x}) h_u(\mathbf{x})-h_t(\mathbf{x}) h_v(\mathbf{x})-h_s(\mathbf{x}) h_u(\mathbf{x})+h_s(\mathbf{x}) h_v(\mathbf{x}).
\]

\begin{theorem}
\label{theo:Owada}
Let $M$ be a closed orientable $\mathcal{C}^2$  Riemannian manifold of dimension $d$ with reach $\tau_M \geq \tau_{\min}$. Let $\X_n = (X_i)_{i=1}^n$ be an i.i.d. sample drawn from a $L$-Lipschitz density $f$ on the manifold where $\mathcal{H}$ is the Hausdorff measure on the manifold.

For $k \in \llbracket 0, d-1 \rrbracket$, denote by $\mu_k$ the measure on $\Delta^+ = \{ (x,y) : 0 \leq x < y  \leq \infty \}$ defined on the rectangles $R_{s, t, u, v} = [s, t) \times [u, v)$ by  
\begin{align*}
\mu_k(R_{s,t,u,v}) = \frac{\int_M f^{k+2} \mathrm{d} \mathcal{H}}{(k+2)!} \int_{(\mathbb{R}^d)^{k+1}} H_{s,t,u,v}(0, y_1, \ldots, y_{k+1}) \mathrm{d}y_{1} \ldots \mathrm{d}y_{k+1},
\end{align*} 
for $0<s \leq t\leq u \leq  v$. For a sequence $r_n$, denote by $\xi_{k,n}$ the re-scaled measure defined by
\begin{align*}
\xi_{k,n}(R_{s,t,u,v}) = \frac{\text{Card}(r_nR_{s,t,u,v}\cap \text{dgm}_k(\C(\X_n/r_n))}{n^{k+2}r_n^{d(k+1)}},
\end{align*}
which counts the number of points of the $k$-th persistence diagram of the rescaled data falling in the rectangle $r_n R_{s, t, u, v}$.  Assume that we are in the sparse divergence regime, i.e. the sequence $r_n$ verifies:
\[ nr_n^d \to 0 \text{ and } n^{k+2}r_n^{d(k+1)} \to \infty.
\]
For $k \leq d-4$, choose $r_n = n^{- \frac{k+2}{2+d(k+1)}}$. Then for $n$ large enough,
\begin{align*}
\sup_{0<s\leq t\leq u \leq v \leq t^+} \mathbb{E} \left[(\xi_{k,n}-\mu_k)(R_{s,t,u,v})^2 \right]  \leq C n^{- \frac{2(k+2)}{2+d(k+1)}}.   
\end{align*}
For $ d-4 \leq  k \leq d$, choose $r_n = n^{-\frac{k+4}{d(k+3)}}$. Then for $n$ large enough,
\begin{align*}
\sup_{0<s\leq t\leq u \leq v \leq t^+} \mathbb{E} \left[(\xi_{k,n}-\mu_k)(R_{s,t,u,v})^2\right]  \leq C n^{- \frac{2}{k+3}},
\end{align*}
where $C$ is a constant that depends only on $k, d, t^{+}, \|f\|_{\infty}, \tau_{\min}$ and $L$.
\end{theorem}

We defer the proof to Section \ref{sec:proof_owada}.

This theorem asserts that asymptotically, the rescaled persistence diagram of the \v{C}ech filtration built on an adequately rescaled point cloud on $\mathbb{R}^d$ converges to a measure $\mu_k$ which depends on $f$ only through $\int_M f^{k+2} \mathrm{d}\mathcal{H}$. Moreover, given two distributions $f_1$ and $f_2$ such that there exists $k \in \llbracket 0, d-1 \rrbracket$ such that $\int_M f_1^{k+2} \mathrm{d}\mathcal{H} \neq \int_M f_2^{k+2} \mathrm{d}\mathcal{H}$, any rectangle $R_{s, t, u, v}$ enables us to distinguish between the two densities $f_1$ and $f_2$ when $n$ is large enough as we make it more explicit in the following corollary. Since this theorem is stated for the rescaled persistence diagram with a sequence $r_n$ that tends to $0$, this is another evidence that points close to the diagonal (even close to the origin) contain information relative to the sampling and should be considered for classification purposes. 

\begin{corollary}
\label{cor:concentration}
Keeping the same notation as above, consider two densities $f_1$ and $f_2$ with Lipschitz constants $L_1$ and $L_2$ such that there exists $k \in \llbracket 0, d-1 \rrbracket$ such that $\int_M f_1^{k+2} \mathrm{d} \mathcal{H} \neq \int_M f_2^{k+2} \mathrm{d} \mathcal{H}$. Let $0 < s \leq t \leq u \leq v$. For $n$ large enough, the number of points in the persistence diagram falling in the rectangle $R_{s, t, u, v}$ identifies the correct sampling density with probability larger or equal than $1-C\frac{n^{-\frac{2(k+2)}{d(k+1)}}}{\left|\int_M (f_1^{k+2}-f_2^{k+2}) \right|^2}$,
 where $C$ is a constant that depends only on $(s, t, u, v), k, d, \|f_1\|_{\infty}, \|f_2\|_{\infty}, \tau_{\min}, L_1$ and $L_2$.

\end{corollary}

The proof of Corollary \ref{cor:concentration} is a straightforward consequence of the Chebyshev's inequality and we defer it to Section \ref{proof:cor_conc}. Deriving a finer concentration inequality is still an open question: indeed, we have only used a bound on the variance of the random variable $\xi_{k, n}$ to use Chebyshev's inequality. While our proof could be adapted to control higher order moments, it would be worth investigating if we could adapt some techniques from the proof of the Theorem 4.5 of \cite{yogeshwaran2017random} to our framework in the sparse regime.

The results derived in Proposition \ref{theo:BD} and Corollary \ref{cor:concentration} both state that the number of sample points $n$ must be large enough to discriminate between the two sampling models with large probability, whether we want to distinguish between manifolds with different homology or different samplings on the same manifold. On the contrary, the results from the previous sections, especially the dependency over $m$ in Proposition \ref{prop:VC} and $\bar{M}_2$ in Theorem \ref{rad_sup} and Proposition \ref{prop:comp} assert that the number of points in the diagram (directly related to the number of sample points) must not be too large in order to obtain a good control of the Rademacher complexity. The number of sample points $n$ acts as a trade-off between the separation of the two classes and a control of the predictive risk.

\subsection{Examples}
\label{sec:illustrations}

\begin{figure}[t]
\begin{center}
\subfigure[$\sigma=0$]{\includegraphics[scale=0.25]{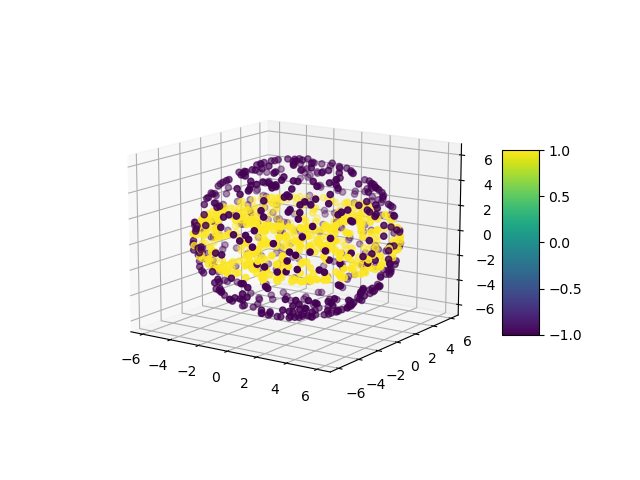}}
\subfigure[$\sigma = 1$]{\includegraphics[scale=0.25]{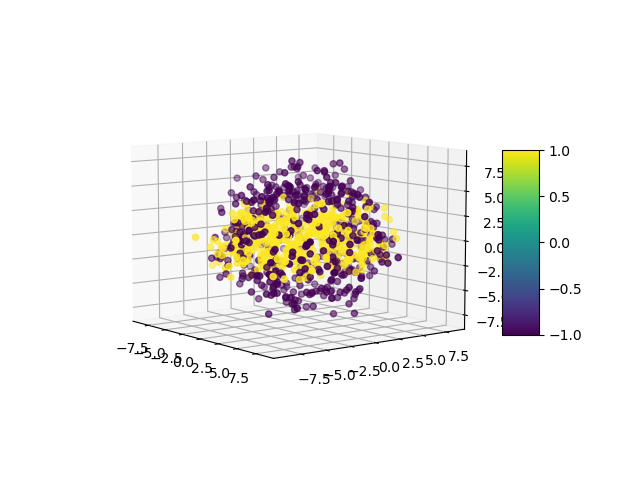}}
\subfigure[$\sigma = 4$]{\includegraphics[scale=0.25]{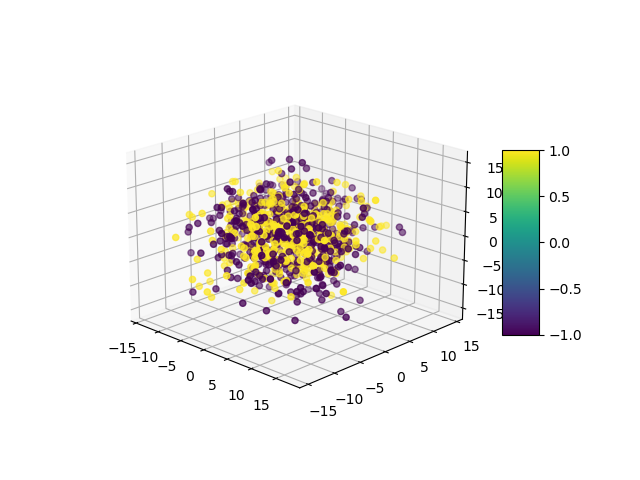}}
\end{center}
\caption{Data to classify. Yellow: torus, purple: sphere.}
\label{data_tor_sph}
\end{figure}

\begin{figure}[t]
\begin{center}
\subfigure[$\sigma=0$]{\includegraphics[scale=0.1]{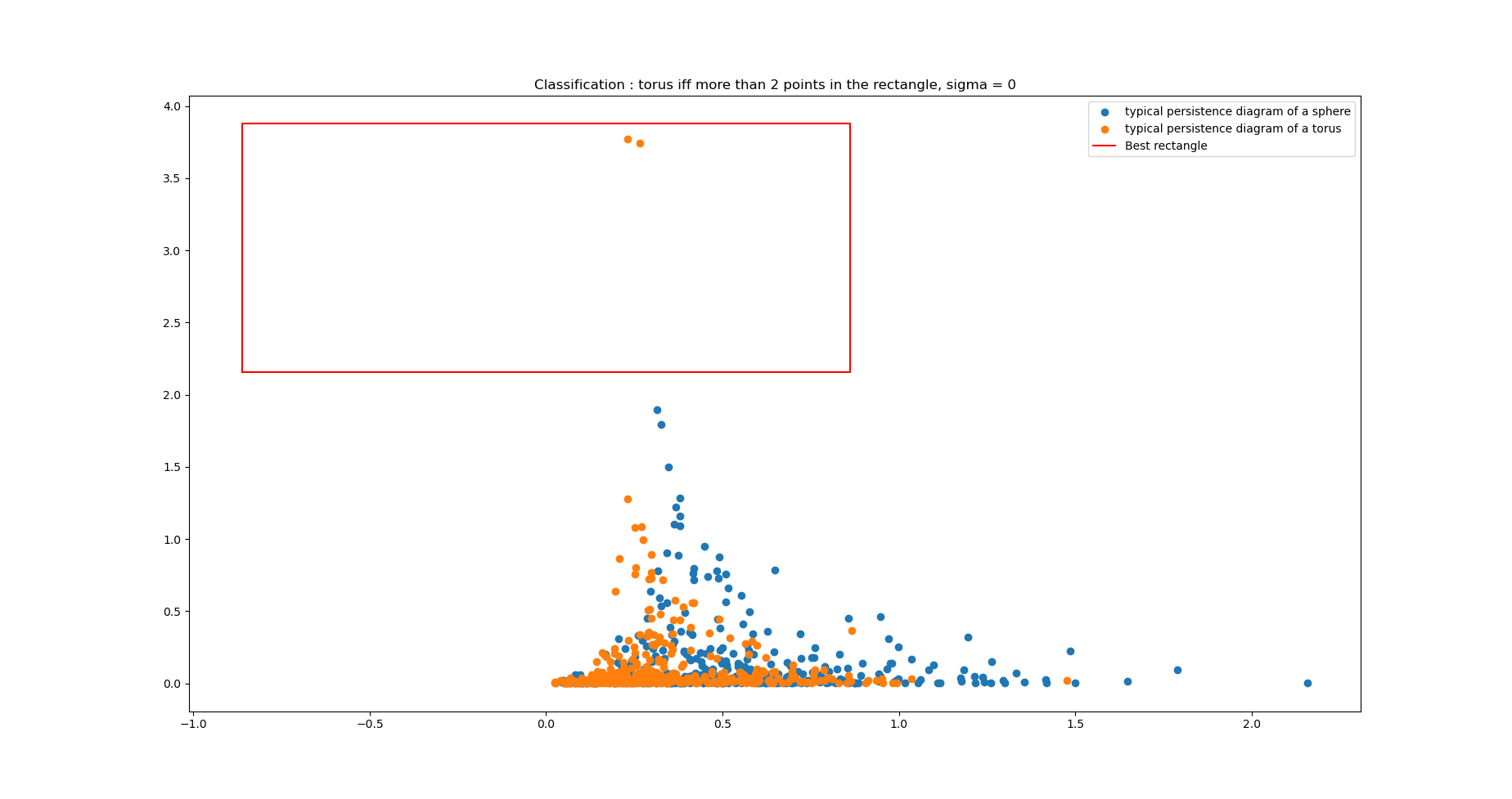}}
\subfigure[$\sigma = 1$]{\includegraphics[scale=0.1]{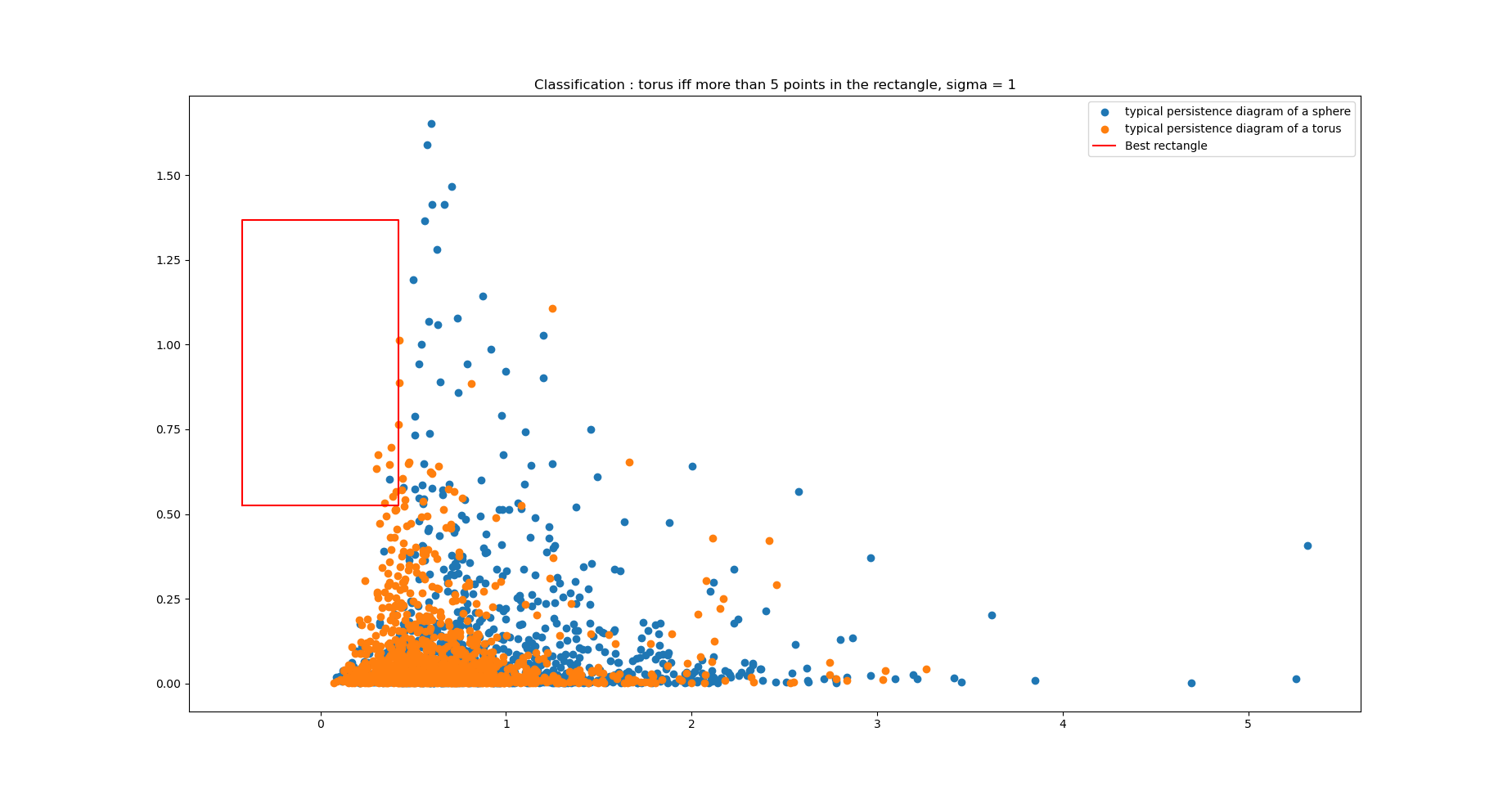}}
\subfigure[$\sigma = 4$]{\includegraphics[scale=0.1]{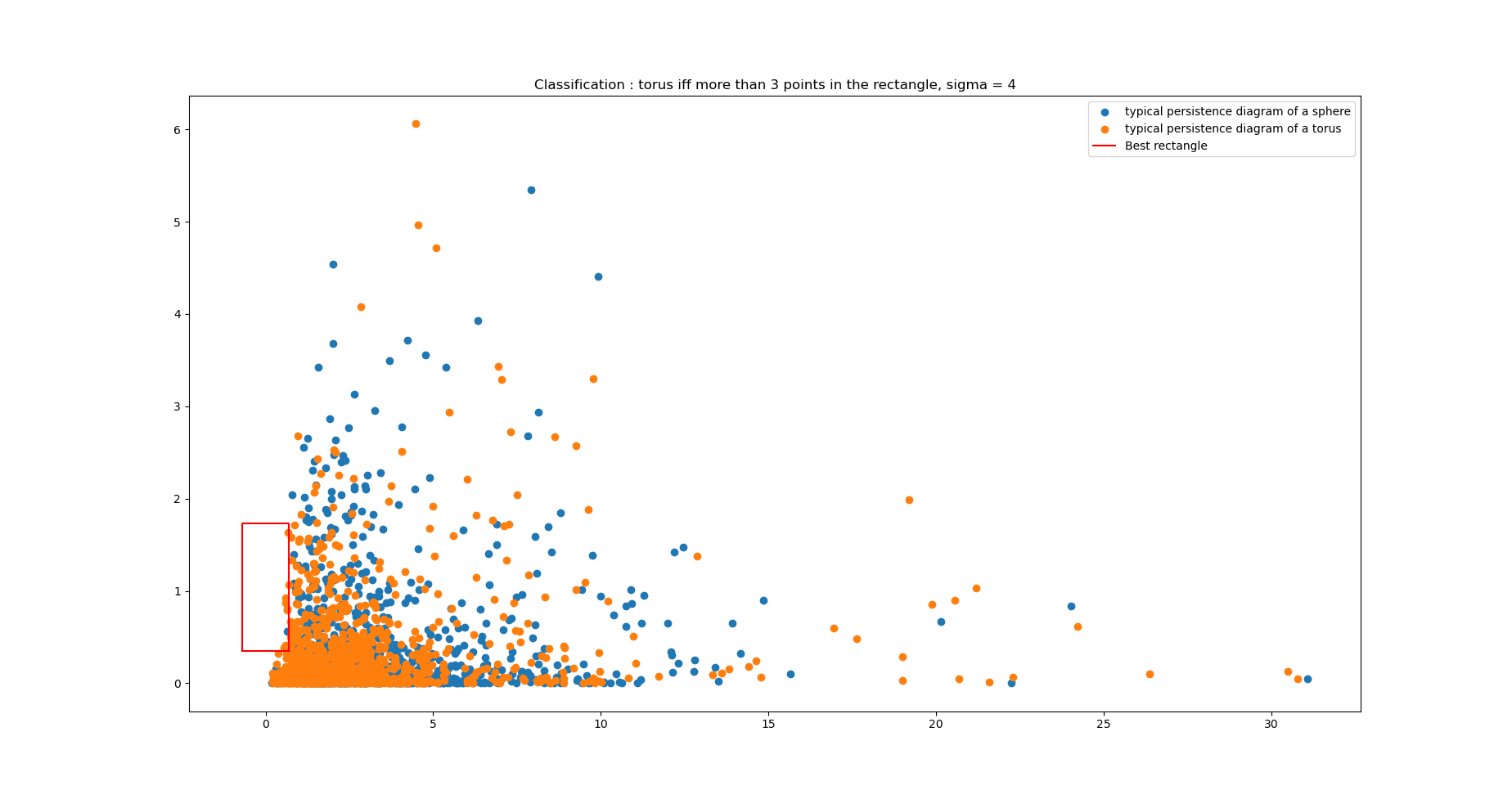}}
\end{center}
\caption{Best rectangle to classify points from a sphere or a torus.}
\label{rect_tor_sph}
\end{figure}

\paragraph*{Sampling on different manifolds}
In this section, we allow ourselves to rotate the diagrams by applying the transformation $(x, y) \mapsto (x, y-x)$, so that all the points lie in the upper-right quadrant, and the diagonal is mapped to the x-axis. In order to illustrate our method, we start by considering $n=500$ points lying on a torus of radii (2, 4) (class $+1$) and a sphere of radius 6 (class $0$). We aim at classifying these data based on the 1-persistence diagram of their \v{C}ech complex. The persistence diagram of the torus is expected to have two high-persistence features. Realizations of the data are shown in Figure \ref{data_tor_sph}. The optimal rectangles minimizing Equation \ref{eq:empir_risk} are displayed in Figure \ref{rect_tor_sph}. 

In the noise-free setting, it is very easy to distinguish between the two classes, both on the raw input and on the persistence diagrams. This corresponds to the framework of Theorem \ref{theo:BD}. If we add a Gaussian noise, it is no longer possible to distinguish which shape is a torus and which is a sphere based on their homology, but it is still possible to distinguish between them because they have different volume measures, by investigating early-born features.

\begin{figure}[t]
\begin{center}
\includegraphics[scale=0.40]{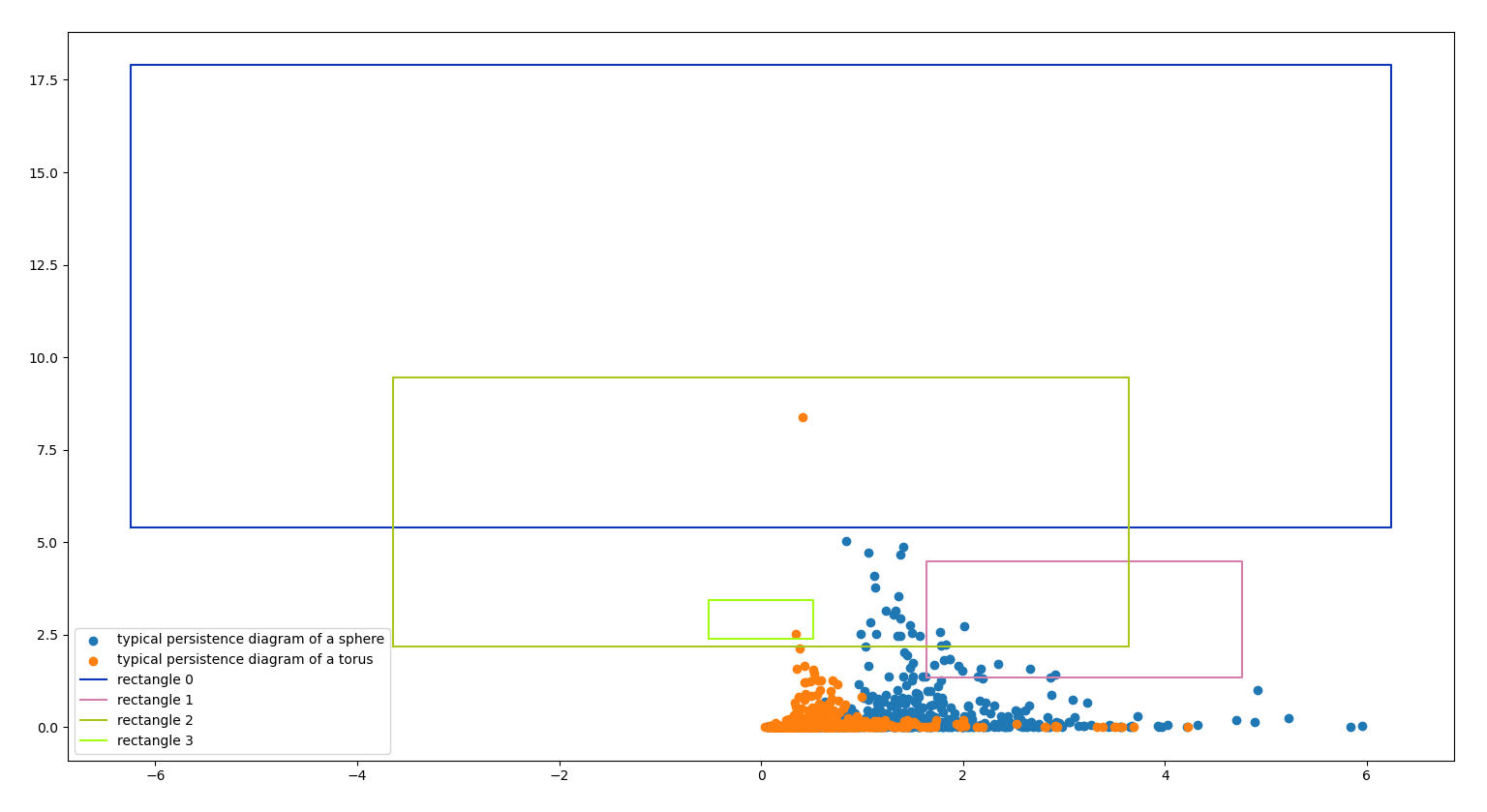}
\end{center}
\caption{Boosting for manifold classification.}
\label{boost_torsph}
\end{figure}

On another experimental set-up, we still aim at distinguishing between point clouds sampled from a torus or a sphere, except that the size of the supporting manifold as well as the number of points are drawn at random. In addition we add a small isotropic noise to the input sample. The illustration of Figure \ref{boost_torsph} shows the first four rectangles of the boosting procedure described at the end of Section \ref{sec:algo}.

The first rectangle aims at discriminating based on the presence of a high-persistence point in the diagram, that would have it classified as a torus (here, there is only one point because of the added noise that makes one of the two features collapse). In this figure, this rectangle alone would suffice to tell the two data apart. However, on other realizations, some of the topological noise from the sphere also belongs to this rectangle. The second rectangle therefore aims at classifying based on the topological noise. Indeed, for points sampled on a torus, cycles will typically be born earlier than on the sphere, and this rectangle aims at detecting late-born cycles, under which circumstances the data will be classified as "sphere". The third rectangle aims at detecting whether there is a significant number of points of high persistence in the topological noise. The fourth one explores if there are features born early, which is a signature of tori. The boosting algorithm aggregates these classifiers and improves the classification performance by up to 10 \% as opposed to considering a single rectangle.

\begin{figure}[t]
\begin{center}
\subfigure[BD between original and noisy samples]{\includegraphics[scale=0.4]{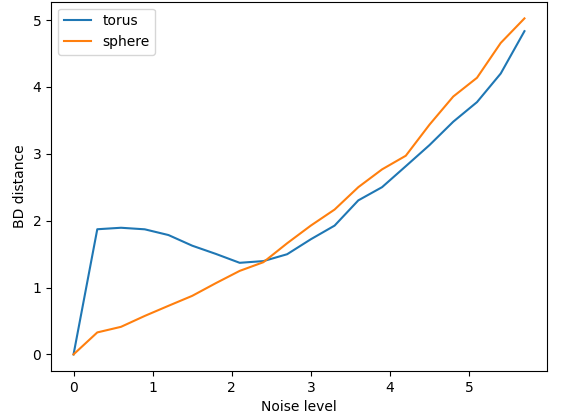}}
\subfigure[Classifier accuracy]{\includegraphics[scale=0.4]{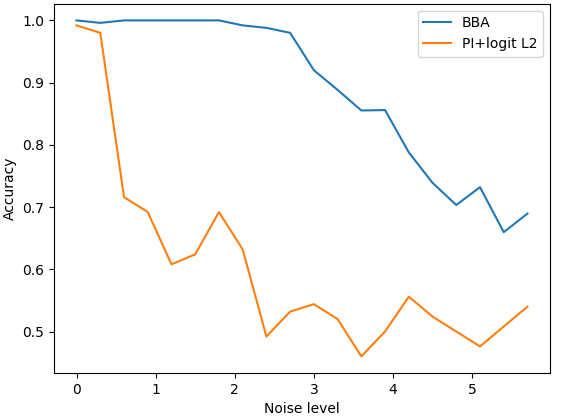}}
\end{center}
\caption{Stability with respect to sampling noise}
\label{fig:stability}
\end{figure}

\paragraph*{Stability analysis}
To assess the stability of our method, we evaluate its performance as the level of isotropic noise $\sigma$ increases. We use the same experimental setup as before: classifying between 250 samples drawn from spheres and 250 samples drawn from tori. Each coordinate of the sample is perturbed by an isotropic noise of magnitude $\sigma$.
We frame this as a supervised learning task, using 250 samples for training and 250 for testing. Classification is based on 1-dimensional persistence diagrams. We compare our approach—an aggregation of 10 balls classifiers (denoted as BBA, or "best balls aggregator")—against an $L^2$-regularized logistic regression classifier applied to persistence images. The results are presented in Figure \ref{fig:stability}.
We first analyze the evolution of the Bottleneck Distance (BD) between a sample on the manifold and its noisy counterpart as a function of $\sigma$. The results, averaged over 100 independent samplings, reveal distinct behaviors: for the sphere, the bottleneck distance increases linearly with $\sigma$, indicating a form of stability. In contrast, even a small amount of noise causes a collapse of the homological features for the torus, resulting in a sharp increase in the bottleneck distance. This instability directly impacts the persistence image classifier, whose accuracy drops significantly.
On the other hand, BBA remains highly robust to such perturbations. This stability can be attributed to its ability to capture fine-grained information within low-persistence features.

\begin{figure}[]
\begin{center}
\subfigure[Samples of point processes]{\includegraphics[scale=0.32]{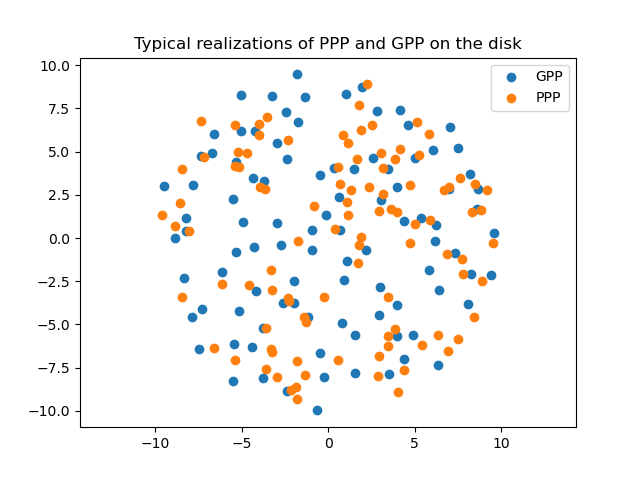}}
\subfigure[Boosting rectangles]{\includegraphics[scale=0.21]{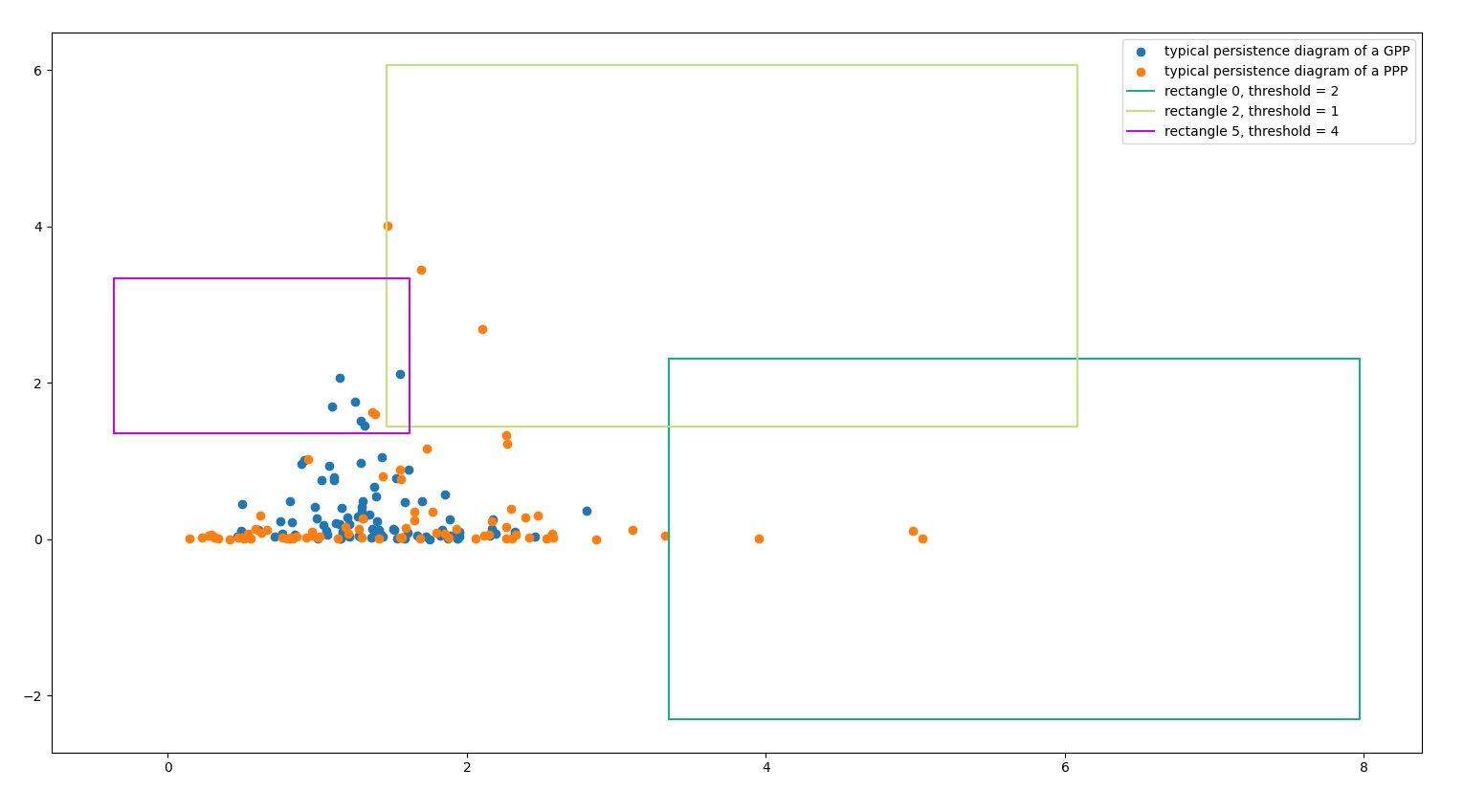}}
\end{center}
\caption{Point processes classification problem.}
\label{ex_PP}
\end{figure}

\paragraph*{Point process classification}
A second experiment conducted is based on the experimental set-up from \cite{obayashi2018persistence}. We sample Poisson (PPP) and Ginibre (GPP) point processes on the disk with 30 points on average, and compute their one-dimensional persistence diagrams. The model has been trained on $400$ processes and tested on $200$. We aggregate the best 10 rectangles with a boosting procedure. We have reached similar classification accuracy (around $94 \%$ in both cases). \cite{obayashi2018persistence} apply a logistic regression to a persistence image transform of the persistence diagrams. When using a $L^1$ penalty, this induces sparsity and highlights a zone of the persistence image useful for discrimination. Our method can be seen as a variation of this where we are free from vectorization and fixed-pixelization when selecting the discriminating support. It is no surprise we obtain similar results on this simple data set. We will actually see in Section \ref{sec:expe} that our method has a better accuracy on real data sets for a comparable running time. We display some of the rectangles found along the boosting optimization in Figure \ref{ex_PP}. A Ginibre point process causes repulsive interactions and points are more evenly spread out, which prevents cycles from dying too early and promotes features with medium-persistence, as we can see on Figure \ref{ex_PP}. In this set-up, there is no "homological signal" to recover, and we only classify based on the topological noise. We only display three rectangles because of overlaps. The first rectangle investigates very late-born cycles of small persistence, which seems to be a characteristic of PPP. Another rectangle looks at features of high-persistence born late, which is once again something promoted by PPP. Note that on this example, this rectangle alone would bring a misclassification. The last rectangle seeks for features of medium persistence born early, and classify as a GPP if there are more than four such features (which is the case here).

\section{Quantitative experiments}
\label{sec:expe}
We compare our method to benchmark datasets in both topological data analysis and point clouds classification. For point cloud data, we consider the \v{C}ech filtration of the point cloud. To simplify the computation, we aggregate the persistence diagrams of different homological dimensions. Although this induces some loss of information as opposed to considering each diagram individually, this is not reflected in practice. Indeed, information from different dimensions typically appears at different scales, see \cite{adler2014crackle}. For all the experiments, we typically perform 10 to 20 boosting iterations where the weak-classifiers are Euclidean balls along with a threshold and where all the parameters are learned by exhaustive search (Algorithm 1 of Section \ref{sec:algo}). At each boosting step, we search for centers of balls among a sub-sample of the $k$-means clusters' centers. When the number of data is somewhat large, we allow ourselves to optimize only over some subset of the available data, taking new data at each boosting step. The $1/\sqrt{N}$ bounds obtained in Section \ref{sec:theory} warrant for the validity of this sub-sampling procedure. When doing multi-class classification, we perform a one-versus-one classification procedure. In the tables below, our method will be denoted by BBA for "best balls aggregator". We have made the code publicly available here.\footnote{ \url{https://github.com/OlympioH/BBA_measures_classification}}

\subsection{Persistence diagrams}
\subsubsection*{Orbit5K}
\begin{figure}[t]
\begin{center}
\subfigure[$\rho=2.5$]{\includegraphics[scale=0.16]{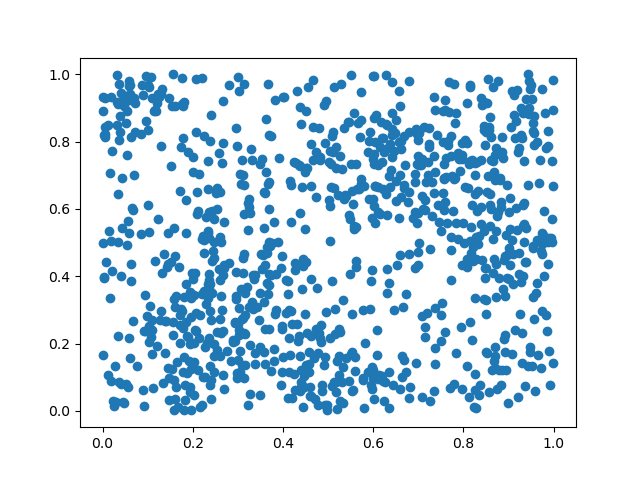}}
\subfigure[$\rho=3.5$]{\includegraphics[scale=0.16]{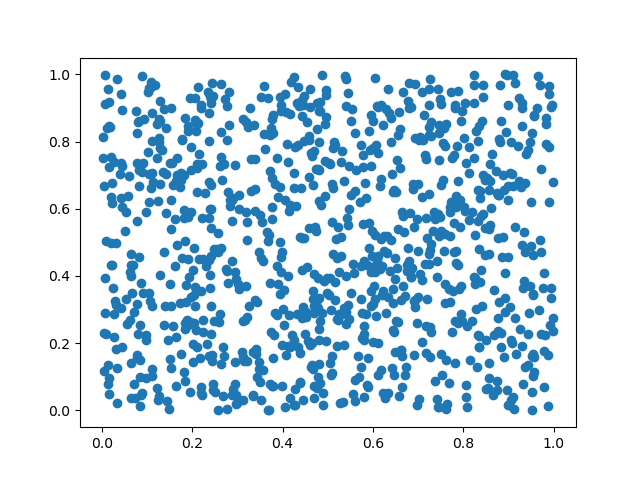}}
\subfigure[$\rho=4.0$]{\includegraphics[scale=0.16]{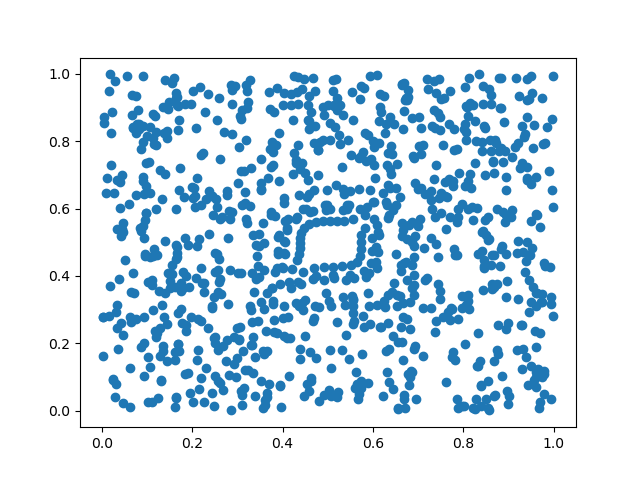}}
\subfigure[$\rho=4.1$]{\includegraphics[scale=0.16]{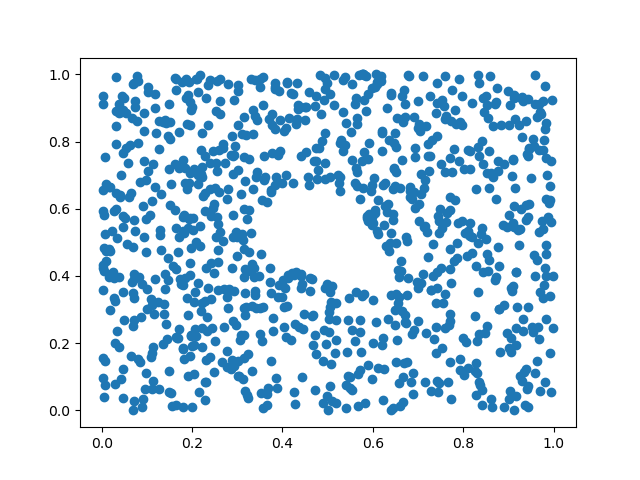}}
\subfigure[$\rho=4.3$]{\includegraphics[scale=0.16]{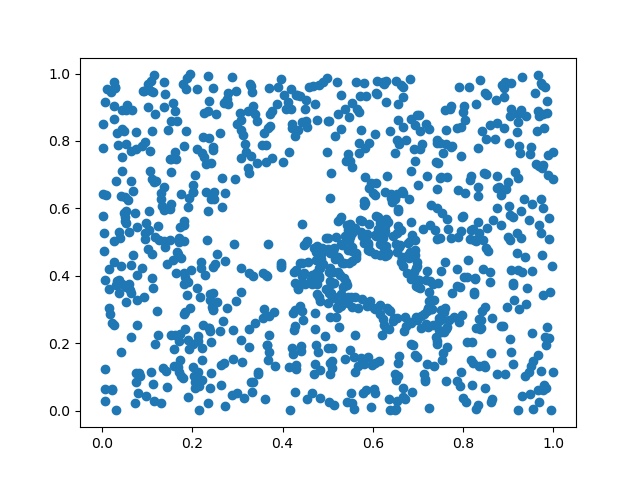}}
\end{center}
\caption{Examples of point clouds from the Orbit5K dataset.}
\label{orbit_PC}
\end{figure}

The dataset ORBIT5K is often used as a standard benchmark for classification methods in TDA. This dataset consists of subsets of size $1000$ of the unit cube $[0, 1]^2$ generated by a dynamical system that depends on a parameter $\rho>0$. To generate a point cloud, a random initial point $(x_0, y_0)$ is chosen uniformly in $[0,1]^2$ and a sequence of points $(x_n, y_n)$ for $n = 0, 1, \ldots , 999$ is generated
recursively by:

\[
\begin{array}{ll}
x_{n+1}=x_{n}+\rho y_{n}\left(1-y_{n}\right) & \bmod \hspace{0.5em} 1 \\
y_{n+1}=y_{n}+\rho x_{n+1}\left(1-x_{n+1}\right) & \bmod \hspace{0.5em} 1.
\end{array}
\]
Given an orbit, we want to predict the value of $\rho$, that can take values in \\ $\{2.5, 3.5, 4.0, 4.1, 4.3 \}$. We display an example for each class in Figure \ref{orbit_PC}; $\rho \in \{4.0, 4.1, 4.3\}$ accounts for difference in topology, while $\rho \in \{2.5, 3.5 \}$ generates samplings with different densities but no particular homological information.

We generate 700 training and 300 testing data for each class. We compare our score with standard classification methods in Table \ref{score:orbit}, where the results are averaged over 10 runs. We compare our scores to four kernel methods on persistence diagrams taken respectively from \cite{reininghaus2015stable}, \cite{kusano2016persistence}, \cite{carriere2017sliced}, \cite{le2018persistence}, and two methods that use a neural network architecture to vectorize persistence diagrams: \cite{carriere2020perslay} and \cite{reinauer2021persformer}. Our accuracy is comparable with kernel methods on persistence diagrams but is somehow lower than that of neural networks.

\begin{table}
\begin{center}
\begin{tabular}{ |c|c|c|c|c|c|c| } 

 \hline
 PSS-K & PWG-K & SW-K & PF-K & Perslay & Persformer & BBA \\ 
 \hline
72.38 $\pm$ 2.4 & 76.63 $\pm$  0.7 & 83.6 $\pm$ 0.9 & 85.9 $\pm$ 0.8 & 87.7 $\pm$ 1.0 & 91.2 $\pm$ 0.8 & 83.3 $\pm$ 0.5 \\ 
 \hline
 
\end{tabular}
\caption{Classification score for the Orbit5K dataset.}
\label{score:orbit}
\end{center}
\end{table}

\subsubsection*{Graph classification}

Another benchmark of experiments in TDA is the classification of graph data. In order to transform graphs into persistence diagrams, we consider the Heat Kernel Signature (HKS) as done by \cite{carriere2020perslay}, for which we recall the construction: for a graph $\mathcal{G} = (V, E)$, the HKS function with diffusion parameter $t$ is defined for each $v \in V$ by
\[hks_t(v) = \sum_{k=1}^{|V|} \exp(-t \lambda_k) \psi_k(v)^2,
\]   
where $\lambda_k$ is the $k$-th eigenvalue of the normalized graph Laplacian and $\psi_k$ the corresponding eigenfunction. We aggregate the points for the 0-dimensional persistence diagram for the sublevel set evolution and the 1-dimensional persistence diagram for the superlevel set evolution. In practice, preliminary experiments suggested that this aggregation induces no loss of information as opposed to considering the two diagrams separately. This corresponds to the \textit{Ord0} and \textit{Ext+} diagrams in extended homology, see \cite{carriere2018structure, carriere2020perslay}. For the experiments, we fixed the value of $t$ to 10, a preliminary study suggested that the diagrams were somehow robust to the choice of this diffusion parameter. The results on standard datasets are provided in Table \ref{score:graphs}. The first five columns are kernel methods or neural networks designed specifically for graph data, P denotes the best method between Persistence image and Persistence landscapes, and MP the best method between multiparameter persistence image, landscape, and kernel (scores reported from \cite{carriere2020multiparameter}). All these persistence-based vectorizations are coupled with a XGBoost classifier to perform the learning task.  We can see that our method clearly outperforms standard vectorizations of persistence diagrams and also multi-persistence descriptors. The accuracy reached is similar to Perslay, \cite{carriere2020perslay} which is a neural network that learns a vector representation of a persistence diagrams and Atol, \cite{royer2021atol} which is another measure learning method. Note that on the biggest dataset \textsc{collab}, our method is clearly outperformed by the other methods, especially Atol. 

\begin{table}
\begin{center}
\begin{tabular}{ |c|c|c|c|c|c|c|c|c|c|c| } 

 \hline
 Dataset & SV & RetGK & FGSD & GCNN & GIN & Perslay & P & MP  & Atol & BBA \\ 
 \hline
Mutag & 88.3 & 90.3 & 92.1 & 86.7 & 89 & 89.8 & 79.2 & 86.1 & 88.3 & 90.4 \\ 
 \hline
 DHFR & 78.4 & 81.5 & - & - & - & 80.3 & 70.9 & 81.7 & 82.7 & 80.5 \\ 
 \hline
  Proteins & 72.6 & 75.8 & 73.4 & 76.3 & 74.8 & 75.9 & 65.4 & 67.5 & 71.4 & 74.7 \\ 
 \hline
     Cox2 & 78.4 & 80.1 & - & - & - & 80.9 & 76.0 & 79.9  & 79.4 & 81.2 \\ 
 \hline
   IMDB-B & 72.9 & 71.9 & 73.6 & 73.1 & 74.3 & 71.2 & 54.0 & 68.7 & 74.8 & 69.4 \\ 
 \hline
    IMDB-M & 50.3 & 47.7 & 52.4 & 50.3 & 52.1 & 48.8 & 36.3 & 46.9 & 47.8 & 46.7 \\ 
 \hline
   COLLAB & - & 81.0 & 80.0 & 79.6 & 80.1 & 76.4 & - &-   & 88.3 & 69.6  \\ 
 \hline

\end{tabular}
\caption{Classification scores for graph data.}
\label{score:graphs}
\end{center}
\end{table}

\subsection{Other datasets}

\subsubsection*{Flow cytometry}
Flow cytometry is a lab test used to analyze cells' characteristics. It is used to perform a medical diagnosis by measuring various biological markers for each cell in the sample. Mathematically, the data are point clouds consisting of tens of thousands of cells living in $\mathbb{R}^D$, where $D$ is the number of biological markers considered. We have trained our model on the Acute Myeloid Leukemia (AML) dataset available here\footnote{ \url{https://flowrepository.org/id/FR-FCM-ZZYA}}. AML is a type of blood cancer that can be detected by performing flow cytometry on the bone marrow cells. The dataset consists of $359$ patients, half of them are used for training and the rest of them for validating the model. For each patient, 7 biological markers are measured across $30000$ cells. We report a $F1$-score of 98.9 \%, while most flow cytometry specific data analysis methods have a score comprised between $95 \%$ and $100 \%$ according to Table 3 from \cite{aghaeepour2013critical}. In addition, our method can lead to qualitative interpretations, since it generates discriminatory zones, and therefore thresholds of activation for biological markers that make a patient sick or healthy. 

\subsubsection*{Time series} 

Another field of applications is the classification of time series. We consider each data as a collection of points by dropping the temporal aspect of the data. We have tried our method on a small sample of data from the University of East Anglia (UEA) archive presented in \cite{bagnall2018uea}. We compare our method against standard classification methods, and report the results from \cite{baldan2021multivariate} in Table \ref{score:TS}

\begin{table}
\begin{center}
\resizebox{\textwidth}{!}{
\begin{tabular}{ |c|c|c|c|c|c|c|c|c|c|c| } 

 \hline
Dataset & dimension & classes & CMFM & LCEM & XGBM & RFM & MLSTM-FCN & ED & DTW & BBA \\ 
 \hline
Heartbeat & 61 & 2 & 76.8 & 76.1 & 69.3 & 80 & 71.4 & 62 & 71.7 &  73.7 \\ 
 \hline
 SCP1 & 6 & 2 & 82 & 83.9 & 82.9 & 82.6 & 86.7 & 77.1 & 77.5 &  77.5 \\ 
 \hline
  SCP2 & 7 & 2 & 48.3 & 55.0 & 48.3 & 47.8 & 52.2 & 48.3 & 53.9 &  56.0 \\ 
 \hline
   Finger Movements & 28 & 2 & 50.1 & 59.0 & 53.0 & 56.0 & 61.0 & 55.0 & 53.0 &  58.0 \\ 
 \hline
   Epilepsy & 3 & 4 & 99.9 & 98.6 & 97.8 & 98.6 & 96.4 & 66.7 & 97.8 &  92.8 \\ 
 \hline
    StandWalkJump & 4 & 3 & 36.3 & 40 & 33.3 & 46.7 & 46.7 & 20 & 33.3 &  46.7 \\ 
 \hline
    Racket Sports & 6 & 4 & 80.9 & 94.1 & 92.8 & 92.1 & 88.2 & 86.8 & 84.2 &  73.7 \\ 
 \hline
 
\end{tabular}}
\caption{Classification scores for multi-dimensional time series dataset.}
\label{score:TS}
\end{center}
\end{table}

Our method competes with the most simple methods for classifying time series, but fails to be state of the art, especially when a high classification score is expected. When there is only little information to be captured (for instance for the datasets StandWalkJump, Finger Movements or SCP2), our method manages to retrieve it. It is to be noted that the comparison cannot be completely fair with respect to methods targeted to specifically deal with time series while we have removed the temporal aspect of the data and only focus on the distribution of the $d$-dimensional data in certain areas of $\mathbb{R}^d$. 

\subsection{Discussion}

\subsubsection*{Computational time}
In order to compare the running time of our method with standard vectorization methods, we consider the problem defined  in Section \ref{sec:illustrations}: we observe points on a torus or a sphere and classify the manifold supporting the point clouds based on their one-dimensional persistence diagrams. We assume the diagrams have been computed in a preliminary step and compare the running time of several methods in Table \ref{tab:time} when classifying over a training set of size 500 or 3000. The average number of points in the one-dimensional persistence diagram is experimentally of the same order as the number of sampled points. For our method, we report the training time for one weak-classifier. In this experiment only one weak-classifier is enough to classify. For the exhaustive search, we have looked for a candidate classifier among a family of balls with $20$ different centers, $10$ different radii and $5$ different thresholds for a total of $2000$ possible classifiers, counting reversed labels. We compare the running times with a Persistence Image of resolution $40 \times 40$ with fixed parameters and we train a logit classifier with a $L^2$ penalty, where the regularization parameter is learned by cross-validation. When the number of sampled points is large enough, most of the computation time is devoted to the vectorization part and only a small fraction of it is dedicated to actually classifying the images. When the number of points is too small, the classification part of the pipeline can take a rather long time. 

The implementation of vectorization methods for persistence diagrams and standard classification algorithms are taken respectively from the Gudhi (\cite{maria2014gudhi}) and Scikit-learn (\cite{pedregosa2011scikit}) libraries. It is likely that our implementation can be improved, leading to a potential computational gain. Nevertheless, an exhaustive search of the best ball-classifier has a comparable running time to that of Lasso-PI + logit L2 which is enough for simple examples. When doing an aggregation procedure of several weak-classifiers, the running time becomes significantly longer but provides a greater accuracy, as noted in Table \ref{score:graphs}. It is also to be noted from Table \ref{tab:time} that our implementation of the optimization of the smoothed objective does not vary much when dealing with large point clouds nor with large datasets, which makes it a preferable candidate for large-scale applications. 

This is backed-up by the timing of some of the graph experiments in Table \ref{tab:time_graph} where we also compare our running times with the Atol method from \cite{chazal2021clustering} for the smallest and biggest graph datasets. For the Atol method, the authors only report the vectorization time without taking into account the training time of a random forest. Note that the average number of nodes and edges in the \textsc{Mutag} dataset are 17.9 and 19.8 while they are of 74.5 and 2457.2 for the \textsc{Collab} dataset, and our method seems to be pretty robust in this increase in scale. We can see that the running time of all methods is comparable. However, in our case, the accuracy of a single weak classifier is quite poor and the BBA method requires about 10 boosting steps to be fully competitive. For small datasets, both in terms of number of points and data, an exhaustive search is highly recommended, also due to the unstable nature of the smooth version which often requires several initializations before finding a relevant classifier.

In practice, the scaling issue of the exhaustive search as dataset size increases is mitigated by only considering a fraction of the dataset at each boosting step iteration. This strategy could be further optimized by introducing adaptive sub-sampling and dynamically adjusting the number of candidate classifiers. A possible approach would be to start with very coarse classifiers trained on the entire dataset and iteratively considering a finer search space.

Finally, note that the method developed here is essentially adapted to binary classification. Although it is possible to extend it to any number of class via a one-versus-one or a one-versus-all classification strategy, this can be computationally prohibitive if the number of classes if high.

\begin{table*}
\begin{center}
\caption{Computational time (in seconds), torus VS sphere.}
\vspace*{0.3cm}

\resizebox{\textwidth}{!}{
\begin{tabular}{|c|c|c|c|c|}
	\hline
Size of the dataset & Size of the point cloud & BBA (smooth) & BBA (exhaustive search) & Lasso-PI + logit-L2 \\
\hline
500 & 100 & 143.0 & 18.7 & 7.4 \\
\hline
3000 & 100 & 139.5 & 97.6 & 151.4 \\
\hline
500 & 500 & 136.8 & 31.8 & 12.0 \\
	\hline	
3000 & 500 & 139.2 & 180.1 & 125.7 \\
\hline
500 & 2000 & 147.4 & 66.6 & 43.1 \\
\hline
3000 & 2000 & 149.7 & 343.9 & 260.1 \\
\hline

\end{tabular}}
\label{tab:time}
\end{center}
\end{table*}

\begin{table*}
\begin{center}
\caption{Computational time (in seconds), graph data}
\vspace*{0.3cm}

\resizebox{\textwidth}{!}{
\begin{tabular}{|c|c|c|c|c|c|}
	\hline
Name of the dataset & Number of data & BBA (smooth) & BBA (exhaustive search) & Lasso-PI + logit-L2 & Atol (vectorization only) \\
\hline
Mutag & 170 & 40.8 & 3.0 & 0.27 & $< 0.1$ \\
\hline
Collab & 4500 & 195.1 & 172.5 & 164.9 & 110 \\
\hline
Collab & 1000 & 175 & 38.6 & 36.2 & - \\
\hline

\end{tabular}}
\label{tab:time_graph}
\end{center}
\end{table*}

\subsubsection*{Take-home message}

The method developed in this article, while being simple and explainable, allows to tackle a wide variety of problems. When used on persistence diagrams, we obtain similar results as kernel methods and manage to come close to some state-of-the-art methods using neural networks on graph data. Our method has a decent performance in terms of accuracy when used on small datasets and a good robustness to sampling noise. When the number of data is larger, the $1/\sqrt{N}$ bounds from Section \ref{sec:theory} justify for training our model on a sub-sample of the dataset and therefore propose a decent accuracy at a mild computational cost. 

In addition, since we locate the areas of the persistence diagram which are the most relevant for classification, this can give information for truncating the simplicial complexes for future applications on the same type of data, and therefore greatly improve the computational time, especially if one is to compute the Rips complex which is known to have a prohibitive number of simplices if untruncated. Due to its simplicity, the natural competitors of our method appear to be standard vectorizations of persistence diagrams coupled with a usual learning algorithm such as logit or random forest. For this type of classifier, we have seen in Table \ref{score:graphs} that our method has a greater accuracy, while having a comparable running time. Beyond persistence diagrams, we have demonstrated that our method offers decent results in a variety of settings and is well suited to dealing with simple data and could be adapted to dealing with large-scale applications.

A natural extension of this work would be to adapt the method for multi-class classification (such as through decision tree ensembles) or to explore its generalisation to unsupervised learning.

\section{Proofs}
\label{sec:proofs}
This section is devoted to the proofs of all theoretical results contained throughout the paper.

\subsection{Proof of Proposition \ref{prop:VC}}
\label{proof:VC}
We denote by $\tilde{\mathcal{F}}$ the class of functions on measures defined by $\tilde{f}(\mu) = [f(x_1), \ldots, f(x_n)]$ for $\mu = \sum_{i=1}^n \delta_{x_i} \in \mathcal{M}_m (\mathcal{X})$ and $f \in \mathcal{F}$.  We denote by $k \mapsto \gamma_{\mathcal{F}} (k)$ the growth function of a hypothesis class $\mathcal{F}$ defined by $ \gamma_{\mathcal{F}}(k) = \underset{x_1, \ldots, x_k}{\sup} \sharp \{(f(x_1), \ldots, f(x_k)| f \in \mathcal{F} \}$. We have that $\gamma_{\tilde{\mathcal{F}}} (N) \leq \gamma_\mathcal{F} (mN)$ since all the measures have at most $m$ points. Using the Sauer-Shelah lemma, we therefore have that $\gamma_{\tilde{\mathcal{F}}} (N)  \leq \left( \frac{emN}{d} \right)^d$ where $d$ is the VC-dimension of $\mathcal{F}$.

Now, consider a set of $d_2$ measures that is shattered by the class $\mathcal{H}$. Using Section 20 of \cite{shalev2014understanding}, we have that for every integer $k$, $\gamma_{\mathcal{H}}(k) \leq \gamma_{\tilde{\mathcal{F}}} (k) \gamma_{\bar{\mathcal{G}}} (k) $ using the observation above Proposition \ref{prop:VC} that $\mathcal{H}$ is a composition class. We therefore have, using Sauer-Shelah lemma again, that:

\[2^{d_2} \leq \gamma_{\mathcal{H}} (d_2) \leq \left( \frac{e m d_2}{d} \right)^d \left(\frac{e d_2}{d^\prime} \right)^{d^\prime}.
\]
Taking the logarithm on both sides and using the same computation as in the proof of Theorem 6 of \cite{sabato2012multi} yields the wanted result.

\subsection{Proof of Lemma \ref{lemma:covering}}

Let $h$ and $g$ be two functions of $\mathcal{F}$.

\[\| \tilde{g}-\tilde{h} \|_{L_p^N} =  \left( \frac{1}{N}  \sum_{i=1}^N (g[\mu_i] - h[\mu_i])^p \right)^{1/p}
\]

\begin{align*}
& \left| \int (h-g) \mathrm{d}\mu_i \right|^p = M_i^p \left| \int (h-g) \mathrm{d}(\mu_i/M_i) \right|^p \\
& \leq M_i^p \int (h-g)^p \mathrm{d} (\mu_i/M_i) \text{ by Jensen inequality}. \\
\end{align*}
Therefore,
\[\|\tilde{h}-\tilde{g}\|_{L_p^N}^p \leq \frac{1}{N} \sum_{i=1}^N M_i^p \int (h-g)^p \mathrm{d} (\mu_i/M_i).
\]
Denoting for each $i$ $w_i = \frac{M_i^p}{\sum_{j=1}^N M_j^p}$, the above inequality writes as :

\begin{align*}
\|\tilde{h}-\tilde{g}\|_{L_p^N}^p &\leq \bar{M}_p \sum_{i=1}^N w_i \int (h-g)^p \mathrm{d} (\mu_i/M_i) \\
&\leq \bar{M}_p \|h-g\|^p_{L_p(\sum_{i=1}^N w_i\mu_i/M_i)}.
\end{align*}
Denoting by $\bar{\mu} = \sum_{i=1}^N w_i\mu_i/M_i$ we have the desired result.

\subsection{Proof of Theorem \ref{rad_sup}}
\label{proof:rad_sup}
By Dudley's chaining theorem, we have that

\begin{align*}
\mathcal{R}_N (\tilde{\mathcal{F}}) & \leq \frac{12}{\sqrt{N}} \int_0^ \infty \sqrt{\ln \mathcal{N}(\tilde{\mathcal{F}}, L_2^n, \varepsilon)} d \varepsilon.
\end{align*}
Remark that
\[
\text{diam} (\tilde{\mathcal{F}}, L_2^n) \leq \frac{1}{\sqrt{N}} \underset{f \in \mathcal{F}} {\sup} \left( \sum_{i=1}^N  \int (f \mathrm{d} \mu_i)^2 \right)^{1/2} \leq \frac{1}{\sqrt{N}} \left( \sum_{i=1}^N M_i^2 \right)^{1/2} \leq \bar{M}_2.
\]
Therefore, we only need to integrate up to $\bar{M}_2$, yielding:

\begin{align*}
\mathcal{R}_N (\tilde{\mathcal{F}}) &\leq \frac{12}{\sqrt{N}} \int_0^{\bar{M}_2} \sqrt{ \ln \mathcal{N} (\mathcal{F}, L_2 (\bar{\mu}), \varepsilon / \bar{M}_2)} d \varepsilon \text{ by the above lemma,} \\
&\leq \frac{12}{\sqrt{N}} \int_0^{\bar{M}_2} \sqrt{K_0 \mathrm{VC}(\mathcal{F}, c\varepsilon/ \bar{M}_2) \ln (2\bar{M}_2/\varepsilon)} d \varepsilon \text{ by Theorem 1 of \cite{mendelson2003entropy},} \\
&\leq \frac{12 \bar{M}_2}{\sqrt{N}} \int_0^1 \sqrt{K_0 \mathrm{VC}(\mathcal{F}, c \varepsilon) \ln (2/\varepsilon)} d\varepsilon \text{by a change of variables,} \\
&\leq \frac{K_1 \bar{M}_2\sqrt{\mathrm{VC}(\mathcal{F})}}{\sqrt{N}} \int_0^1 \sqrt{\ln(2 / \varepsilon)} d \varepsilon. \\
&
\end{align*}
Here, $K_0$ and $K_1$ are universal constants. Including the integral in the multiplicative constant term gives the wanted result.

\subsection{Proof of Theorem \ref{rad_inf}}

By Sudakov minoration principle, there exists a constant $C$ such that for all $\varepsilon >0$, 

\[\frac{C \varepsilon}{\sqrt{N}} \sqrt{\ln \mathcal{N}(\tilde{\mathcal{F}}, \varepsilon, L_2^N)} \leq  G_N (\tilde{\mathcal{F}}),
\]
where $G_N$ stands for the Gaussian complexity.

Classical equivalence between covering and packing numbers yields 

\[\frac{C \varepsilon}{\sqrt{N}} \sqrt{\ln \mathcal{M}(\tilde{\mathcal{F}}, 2\varepsilon, L_2^N)} \leq  G_N (\tilde{\mathcal{F}}).
\]
If all the $\mu_i$ are of the form $M_i \delta_{x_i}$ for $(x_i)_{i=1, \ldots, N} \in \mathcal{X}^N$, we have for two functions $g$ and $h$ in $\mathcal{F}$ that
\[\|\tilde{g}-\tilde{h}\|_{L_2(\mu_1^N)} = \frac{1}{N}\sqrt{\sum_{i=1}^N(g[\mu_i] - h[\mu_i])^2} = \frac{1}{N}\sqrt{\sum_{i=1}^N M_i^2(g(x_i)-h(x_i))^2}=\bar{M}_2 \|g-h\|_{L_2(x_1^N, w)},
\]
for the $L_2$-norm with weights $w_i = \frac{M_i^2}{\sum_{j=1}^N M_i^2}$.

When looking on the supremum over all measures, we can therefore lower bound the packing number:

\begin{align*}
\underset{(\mu_1, \ldots, \mu_N) \in \mathcal{C}_{\bar{M}_2}^N}{\sup} G_N (\tilde{\mathcal{F}}) & \geq \frac{C\varepsilon}{\sqrt{N}}\ln \sqrt{\underset{(\mu_1, \ldots, \mu_N) \in \mathcal{C}_{\bar{M}_2}^N}{\sup} \mathcal{M}(\tilde{\mathcal{F}}, 2\varepsilon, L_2(\mu_1^N))} \\ 
& \geq \frac{C\varepsilon}{\sqrt{N}}\ln \sqrt{\underset{x_1, \ldots, x_N}{\sup} \mathcal{M}(\mathcal{F}, 2\varepsilon/\bar{M}_2, L_2(x_1^N, w))}.
\end{align*}

In particular, by taking $x_1, \ldots x_N$ that are $2\varepsilon/\bar{M}_2$-shattered by $\mathcal{F}$ if $N \leq \mathrm{VC}(\mathcal{F}, 2 \varepsilon/ \bar{M}_2)$ along with uniform weights, Proposition $1.4$ from \cite{talagrand2003vapnik} states that the logarithm of the packing number dominates the fat-shattering function. If $N >\mathrm{VC}(\mathcal{F}, 2 \varepsilon/ \bar{M}_2)$, the same result simply follows by considering the uniform measure on $\mathrm{VC}(\mathcal{F}, 2 \varepsilon/ \bar{M}_2)$ of the $N$ points and setting weight $0$ to the others.

This together with the equivalence between Gaussian and Rademacher complexities yields that for all $\varepsilon >0$,

\[ K^\prime  \frac{\varepsilon}{\sqrt{N} \ln(N)}  \sqrt{\mathrm{VC}(\mathcal{F}, 4\varepsilon/\bar{M}_2)}\leq \underset{(\mu_1, \ldots, \mu_N) \in \mathcal{C}_{\bar{M}_2}^N}{\sup} \mathcal{R}_N \left( \tilde{\mathcal{F}}|\mu_1, \ldots, \mu_N \right).
\]
In particular, taking $\varepsilon = \bar{M}_2/8$ gives the wanted result, by noticing that for the classification problem, i.e. labels in $\{0, 1\}$, we have that $\mathrm{VC}(\mathcal{F}, 1/2) = \mathrm{VC}(\mathcal{F})$.

\subsection{Proof of Proposition \ref{prop:comp}}
\label{proof:comp}

Let $\bar{\mu} = (\mu_1, \ldots, \mu_N)$ be a sample of $N$ measures. In Theorem 2 of \cite{maurer2016chain}, the authors establish a chain-rule to control the Gaussian complexity for the composition of function classes. This result implies that there exist two constants $C_1$ and $C_2$ such that for any $f_0 \in \mathcal{F}$,

\[ G_N(\mathcal{H}) \leq C_1 L G_N(\tilde{\mathcal{F}}) +\frac{1}{N}
C_2 \text{Diam} (\tilde{\mathcal{F}}(\bar{\mu})) \mathbf{R}(\mathcal{G}) + G_N ( \mathcal{G}(f_0)).
\]

\[\text{ where } \mathbf{R}(\mathcal{G}) = \sup _{\mathbf{x}, \mathbf{x}^{\prime} \in \mathbb{R}, \mathbf{x} \neq \mathbf{x}^{\prime}} \mathbb{E}_{\gamma} \sup _{\psi \in \mathcal{\mathcal{H}}} \frac{(\psi(\mathbf{x})-\psi(\mathbf{x^\prime))} \gamma}{|\mathbf{x}-\mathbf{x}^{\prime}|},
\] 
and where $\gamma \sim \mathcal{N} (0, 1)$.

We wish to successively bound each of the three terms on the right hand side. The classical equivalence between Gaussian and Rademacher complexities together with Theorem \ref{rad_sup} permits to control the first term:

\[ G_N(\mathcal{H}) \leq \frac{C_1 L \bar{M}_2 \sqrt{\mathrm{VC}(\mathcal{F})} \sqrt{\log(N)}}{\sqrt{N}} + \frac{1}{N}
C_2 \text{Diam} (\tilde{\mathcal{F}}(\bar{\mu})) \mathbf{R}(\mathcal{G}) +  G_N ( \mathcal{G}(f_0)).
\]

Analogously to the proof of Theorem \ref{rad_sup}, we can simply bound the diameter by $\text{Diam}(\tilde{\mathcal{F}}(\bar{\mu})) \leq \sqrt{N} \bar{M}_2$, since all the functions from $\mathcal{F}$ are bounded by 1.

As for the third term, taking $f_0 = 0$, we have that

\begin{align*}
G_N(\mathcal{G}(\tilde{f}_0) &= \mathbb{E}_\gamma \left[\underset{\psi \in \mathcal{H}}{\sup} \langle \gamma, (\psi(0), \ldots, \psi(0) \rangle \right] \\
 &\leq  \mathbb{E}_\gamma \left[ \left| \sum_{i=1}^N \gamma_i \right| \times \underset{\psi \in \mathcal{H}}{\sup} |\psi(0)| \right] \\
 &\leq \sqrt{N} \underset{\psi \in \mathcal{H}}{\sup} |\psi(0)|,
\end{align*}
where the last inequality follows from the fact that the $\gamma_i$ are standard independent normal variables.

\subsection{Proof of Theorem \ref{theo:BD}}
\label{proof:BD}

Assume that we observe a $n$-sample from $\mathbb{M}_1$. Corollary 3 of \cite{chazal2014convergence} states that for every $\varepsilon >0$, 

\[\mathbb{P}[ d_B (\mathrm{dgm}(\C(\mathbb{M}_1)), \mathrm{dgm}(\C(\hat{X}_n))) \geq \varepsilon ] \leq \frac{2^b}{a \varepsilon^b} \exp(-na\varepsilon^b).
\]
Let $\delta >0$, and take $\varepsilon = K/2$. The above formula yields that for $n \geq \frac{2^b}{aK^b} \log \left(\frac{4^b}{aK^b \delta} \right)$, we have with probability larger than $1-\delta$ that

\begin{equation}
\label{cvgKsur2}
d_B (\mathrm{dgm}(\C(\mathbb{M}_1)), \mathrm{dgm}(\C(\hat{X}_n))) \leq K/2. 
\end{equation}

By triangle inequality for the distance $d_B$ and using the hypothesis that the persistence diagrams of the two metric spaces are away from at least $K$ for the bottleneck distance, we necessarily have that

\begin{equation}
\label{awayKsur2}
d_B (\mathrm{dgm}(\C(\mathbb{M}_2)), \mathrm{dgm}(\C(\hat{X}_n))) > K/2.
\end{equation}

By assumption, for $i=1,2, \mathrm{dgm}(\C(\mathbb{M}_i))$ has no point at a distance less than $K$ from the diagonal. We can now distinguish two cases :
\begin{itemize}
\item $\mathrm{dgm} (\C(\mathbb{M}_1))$ and $\mathrm{dgm} (\C(\mathbb{M}_2))$ have the same number of points $m$ (all these points are at least away from $K$ to the diagonal). Under these circumstances, $\mathrm{dgm}(\C(\hat{X}_n))$ also has $m$ points above $K/2$. If it had more, it would mean that one of this point should be matched with the diagonal, and therefore yields a contradiction with  (\ref{cvgKsur2}). Consider squares of size $K/2$ centered on the points of $\mathrm{dgm}(\C(\mathbb{M}_1))$ and $\mathrm{dgm}(\C(\mathbb{M}_2))$. If they all contain the same number of points from $\mathrm{dgm}(\C(\hat{X}_n))$, we have a contradiction with (\ref{awayKsur2}). It therefore means that there is a rectangle that can select the right model.
\item If they do not have the same number of points, necessarily by (\ref{cvgKsur2}), $\mathrm{dgm}(\C(\hat{X}_n))$ must have the same number of points as $\mathrm{dgm}(\C(\mathbb{M}_1))$ above $K/2$ and do not have the same number of points as $\mathrm{dgm}(\C(\mathbb{M}_2))$. Counting the number of points in the (infinite but truncatable) rectangle $ \{ (p,q) | |q-p| > K/2 \}$ is therefore enough to classify between the two metric spaces.
\end{itemize}

\subsection{Proof of Theorem \ref{theo:Owada}}

\label{sec:proof_owada}

We define the persistent Betti number $\beta_{k,n} (a,b)$ as the number of $k$-holes of $\C(r_n^{-1} \X_n, r)$ that persist between $r=a$ and $r=b$. It corresponds to the number of points in the persistence diagram that falls in the upper-left quadrant having an angle at the point $(a,b)$.

First note that $\text{Card}(r_nR_{s,t,u,v}) = \beta_{k,n}(t,u) - \beta_{k,n}(t,v) - \beta_{k,n}(s,u) + \beta_{k,n}(s,v)$. As in \cite{Owada21}, we denote by
\begin{align*}
h_r(x_{1}, \ldots x_{k+2}) = \1_{\bigcap_{j_0=1}^k \{\bigcap_{j \neq j_0} \B(x_j,r/2) \neq \emptyset \}} - \1_{\bigcap_{j=1}^{k+2} \B(x_j,r/2) \neq \emptyset},
\end{align*}
and by 
\begin{align*}
G_{k,n}(s,t) = \sum_{\Y \subset \X_n, |\Y| = k+2} h_{r_ns}(\Y) h_{r_nt}(\Y), 
\end{align*}
so that, according to \cite[Lemma 4.1]{Owada21} 
\begin{align}\label{eq:encadrement_bettipers}
G_{k,n}(s,t) - \binom{k+3}{k+2} L_{r_nt} \leq \beta_{k,n}(s,t) \leq G_{k,n}(s,t) + \binom{k+3}{k+1} L_{r_nt},
\end{align}
where 
\begin{align*}
L_{r_nt} = \sum_{\mathcal{Y} \subset \X_{n}, |\mathcal{Y}| = k+3} \1_{\C(\mathcal{Y},r_nt)\mbox{ is connected} }.
\end{align*}
In what follows we prove bounds for $\beta_{k,n}(s,t)$. The bound on $\text{Card}(r_nR_{s,t,u,v})$ easily follows.
\subsection*{Upper-bound of the bias}

\begin{align*}
\E(G_{k,n}(s,t)) & = \binom{n}{k+2} \int_M f(x_1) \mathrm{d} \mathcal{H}(x_1) \int_{M^{k+1}} g_{s,t} \left ( \frac{x_1}{r_n}, \ldots, \frac{x_{k+2}}{r_n} \right ) \prod_{j=2}^{k+1} f(x_j) \mathrm{d}\mathcal{H}(x_j) \\
& = \binom{n}{k+2}\int_M f(x_1) \mathrm{d}\mathcal{H} (x_1) I_{x_1},
\end{align*}
where $g_{s,t} = h_s h_t$. Now, for a fixed $x_1 \in M$, we note that $g_{s,t}$ is non-zero implies $(x_2, \ldots x_{k+1}) \in \B(x_1, r_n(k+2)t^+)^{k}$ (recall that $t \leq t^+$). Denoting by $\tilde{M}_n = h_n(M)$, with $h_n: u \mapsto \frac{u-x_1}{r_n}$,  and using \citet[Theorem 3.1]{Federer59} leads to the change of variable
\begin{align*}
I_{x_1} &  := \int_{M^{k+1}} g_{s,t} \left ( \frac{x_1}{r_n}, \ldots, \frac{x_{k+2}}{r_n} \right ) \prod_{j=2}^{k+1} f(x_j) \mathrm{d}\mathcal{H}(x_j) \\
& = r_n^{d(k+1)} \int_{(\tilde{M}_n)^{k+1}} g_{s,t} \left ( 0,y_1, \ldots, y_{k+1} \right ) \1_{\B(0,(k+2)t^+)^{k+1}}(y_1, \ldots, y_{k+1}) \prod_{j=1}^{k+1} f(x_1 + r_ny_j) \mathrm{d}\mathcal{H}(y_j).
\end{align*}

Note that $0 \in \tilde{M}_n$, and that $\tilde{M}_n$ has a reach $\tilde{\tau}=\tau/r_n \rightarrow + \infty$. With a slight abuse of notation, we identify $T_0 \tilde{M}_n$ with $\R^d$, and denote by $J_v$ the Jacobian of the exponential map $\exp_0: B_{\R^d}(0, (k+2)t^+) \rightarrow \tilde{M}_n$ at point $v$ (note that $\exp_0$ is well defined for $n$ large enough so that $\tilde{\tau} \geq 4(k+2)t^+$, see, for instance \cite[Lemma 1]{Aamari19}). Using \cite[Theorem 3.1]{Federer59} again yields for the change of variable $y_j = \exp_0 (v_j)$ that
\begin{align*}
I_{x_1} = r_n^{d(k+1)} \int_{(R^d)^{k+1}}  g_{s,t} \left ( 0,y_1, \ldots, y_{k+1} \right ) \1_{y_{1}, \ldots, y_{k+1}\in \B_{\tilde{M}_n}(0,(k+2)t^+)^{k+1}} \prod_{j=1}^{k+1} J_{v_j} f(x_1 + r_ny_j) dv_{1} \ldots dv_{k+1}.
\end{align*}
 According to \cite[Lemma 1]{Aamari19}, whenever $y_j \in \B_{\tilde{M}_n}(0,(k+2)t^+)$, we have $\|y_j - v_j \| \leq C ((k+2)t^+)^2r_n/\tau_{\min}$, and $\|\d_{v_i} \exp_0 - I_d \|_{op} \leq \frac{5}{4\tilde{\tau}_n}=\frac{5r_n}{4 \tau} \leq \frac{5r_n}{4 \tau_{\min}}$,
 so that 
\begin{align*}
|J_{v_j}-1| \leq C_d \frac{r_n}{\tau_{\min}},
\end{align*}
and therefore, $\left| J_{v_j}-1 \right| \leq 1$ for $n$ large enough. We deduce that
\begin{align*}
\left |\prod_{j=1}^{k+1} J_{v_j} f(x_1 + r_ny_j)  - f(x_1)^{k+1} \right | & \leq \left |\prod_{j=1}^{k+1} J_{v_j} f(x_1 + r_ny_j) - \prod_{j=1}^{k+1} J_{v_j} f(x_1) \right | \\
& + \quad \left | \prod_{j=1}^{k+1} J_{v_j} f(x_1) - f(x_1)^{k+1} \right | \\
& \leq C_d^{k+1} (k+1)L \|f\|_{\infty}^{k}r_n + (k+1)\|f\|_{\infty}^{k+1} C_d^{k+1} \frac{r_n}{\tau_{\min}} \\
& \leq C_d^{k+1}(k+1)\|f\|_{\infty}^{k} \left( L \vee \frac{\|f\|_{\infty}}{\tau_{\min}} \right) r_n. 
\end{align*}

Denoting by 
\[
I'_{x_1} = r_n^{d(k+1)} f(x_1)^{k+1}\int_{(R^d)^{k+1}}  g_{s,t} \left( 0,y_1, \ldots, y_{k+1} \right) \1_{y_{1}, \ldots, y_{k+1} \in \B_{\tilde{M}_n}(0,(k+2)t^+)^{k+1}} \prod_{j=1}^{k+1} \mathrm{d}v_1 \ldots \mathrm{d}v_{k+1},
\] we deduce that
\begin{align*}
|I'_{x_1}-I_{x_1}| & \leq  (2(k+2)t^+)^{d(k+1)} C_d^{k+1}(k+1)\|f\|_{\infty}^{k+1} \left( L \vee \frac{1}{\tau_{\min}} \right)r_n^{d(k+1) +1} \\
& \leq C_{d,k}(t^+)^{d(k+1)}\|f\|_{\infty}^{k} \left( L \vee \frac{\|f\|_{\infty}}{\tau_{\min}} \right)r_n^{d(k+1)+1}.
\end{align*} 
Next, note that   
\begin{align*}
g_{s,t} \left ( 0,y_1, \ldots, y_{k+1} \right ) \neq g_{s,t} \left ( 0,v_1, \ldots, v_{k+1} \right ) \Rightarrow (v_1, \ldots, v_{k+1}) \in V_1,
\end{align*}
where 
\begin{align*}
V_1 & = \left \{ (v_1, \ldots, v_{k+1}) \in \B(0, 2(k+2)t^+)^{k+1} \mid \exists i \neq j ;| \|v_i-v_j\| -s | \leq C ((k+2)t^+)^2r_n/\tau_{\min} \right . \\
& \quad  \mbox{or} \quad \left . | \|v_i-v_j\| -t | \leq C ((k+2)t^+)^2r_n/\tau_{\min} \right \}.
\end{align*}

We deduce that
\begin{align*}
& \left | I'_{x_1}(A) - r_n^{d(k+1)}f(x_1)^{k+1} \int_{(R^d)^{k+1}}  g_{s,t} \left ( 0,v_1, \ldots, v_{k+1} \right ) \1_{v_1, \ldots v_{k+1} \in \B_{\tilde{M}_n}(0,2(k+2)t^+)^{k+1}} \mathrm{d}v_1 \ldots \mathrm{d}v_{k+1} \right | \\
& \quad  \leq r_n^{d(k+1)} \|f\|_{\infty}^{k+1} \int_{(R^d)^{k+1}} \1_{V_1}(v_1, \ldots, v_{k+1}) \mathrm{d}v_1, \ldots,  \mathrm{d}v_{k+1} \\
& \quad \leq r_n^{d(k+1)} \|f\|_{\infty}^{k+1} 2 \binom{k+1}{2} C_d ((2(k+2)t^+)^{kd}((2(k+2)t^+)^{d+1}\frac{r_n}{\tau_{\min}} \\
& \quad \leq r_n^{d(k+1)} C_{d,k} \|f\|_{\infty}^k (t^+)^{(k+1)d} \frac{\|f\|_{\infty} t^+ r_n}{\tau_{\min}}.
\end{align*}
The triangle inequality gives
\begin{align*}
\left | \frac{\E (G_{k,n}(s,t))}{\binom{n}{k+2} r_n^{d(k+1)}} - A_k(s,t) \right | \leq C_{d,k,t^+,\|f\|_{\infty}} \left( L \vee \frac{\|f\|_{\infty}}{\tau_{\min}}\right)r_n,
\end{align*}
where $A_{k}(s,t) = \left ( \int_M f^{k+1}(u) \H^d(du) \right ) \int_{(R^d)^{k+1}}  g_{s,t} \left ( 0,v_1, \ldots, v_{k+1} \right )  \mathrm{d}v_1 \ldots \mathrm{d}v_{k+1}$.

Next, we have to bound the higher order term $\E(L_{r_n,t})$ dealing with the subsets of size $k+3$.
To do so, write
\begin{align*}
\E(L_{r_nt}) & = \binom{n}{k+3} \int_{M^{k+3}} \1_{\C(x_1, \ldots, x_{k+3},r_nt)\mbox{ is connected} } \prod_{i=1}^{k+3} f(x_i) \mathrm{d}\mathcal{H}(x_i) \\
& = \binom{n}{k+3} \int_M f(x_1)\mathrm{d}\mathcal{H}(x_1) \int_{M^{k+2}} \1_{\C(x_1, \ldots, x_{k+3},r_nt)\mbox{ is connected} }  \\
 & \qquad \qquad \qquad \qquad \times \1_{x_2, \ldots x_{k+3} \in \B(x_1, (k+3)r_n t)^{k+2}} \prod_{i=2}^{k+3} f(x_i) \mathrm{d}\mathcal{H}(x_i) \\
& \leq C_d \|f\|_{\infty}^{k+2} \binom{n}{k+3}  \int_{M} f(x_1) ((k+3)r_nt^+)^{d(k+2)} \mathrm{d}\mathcal{H}(x_1) \\
& \leq C_{d,k,t^+,\|f\|_{\infty}} \binom{n}{k+3} r_n^{d(k+2)},   
\end{align*}
according to \cite[Lemma B.7]{Aamari19}, since $(k+3)r_nt^+ \leq \tau_{\min}/4$ for $n$ large enough. Thus, 
\begin{align*}
\binom{n}{k+2}^{-1} r_n^{-d(k+1)} \E(L_{r_n,t}) \leq C_{d,k,t^+,\|f\|_{\infty}} nr_n^{d}.
\end{align*}

\subsection*{Upper-bound of the variance}

Let us denote by 

\begin{align*}
U_n = \frac{1}{\binom{n}{m}} \sum_{I \subset \X_n, |I|=m} g_{s,t}(I),
\end{align*}
where $m=k+2$, and, for $j =1, \hdots, m-1$, 
\begin{align*}
g_j(x_1, \ldots x_j) = \E \left ( g_{s,t}(x_1, \ldots, x_j, X_{j+1}, \ldots, X_m \right ).
\end{align*}
We remark that $g_m = g_{s,t}$. Noting that $U_n$ is a U-statistics of order $m$, Hoeffding's decomposition (see, e.g., \cite[Theorem 3]{Lee90}) yields that
\begin{align}\label{eq:var_hoeffding}
\Var(U_n) = {\binom{n}{m}}^{-1} \sum_{j=1}^m \binom{m}{j} \binom{n-m}{m-j} \Var(g_j).
\end{align}
Proceeding as for the bound on $\E(L_{r_nt})$, we may write
\begin{align*}
|g_j(x_1, \ldots x_j)| \leq C_{d,m,t^+,\|f\|_{\infty}} r_n^{d(m-j)} \1_{\C(x_1, \ldots x_j) \mbox{is connected}}, 
\end{align*}
so that 
\begin{align*}
\Var(g_j(X_1, \ldots, X_j)) \leq \E(g_j^2(X_1, \ldots, X_j) \leq C_{d,m,t^+,\|f\|_{\infty}} r_n^{2d(m-j) + (j-1)d},
\end{align*}
for $j = 1, \hdots, m-1$. As well, 
\begin{align*}
\Var(g_m(X_1, \ldots, X_m)) \leq \E(g_m^2(X_1, \ldots, X_m)) \leq C_{d,m,t^+,\|f\|_{\infty}} r_n^{(m-1)d}.
\end{align*}
Plugging these inequalities into \eqref{eq:var_hoeffding} leads to
\begin{align*}
\Var(U_n)& \leq {\binom{n}{m}}^{-1} C_{d,m,t^+,\|f\|_{\infty}} \sum_{j=1}^m \binom{m}{j} \binom{n-m}{m-j} r_n^{2d(m-j) + (j-1)d} \\ 
& \leq C_{d,m,t^+,\|f\|_{\infty}}r_n^{d(2m-1)} \sum_{j=1}^m \frac{\binom{n-m}{m-j}}{\binom{n}{m}}r_n^{-dj} \\
& \leq C_{d,m,t^+,\|f\|_{\infty}}r_n^{d(2m-1)} \sum_{j=1}^m \frac{1}{(nr_n^d)^j} \\
& \leq C_{d,m,t^+,\|f\|_{\infty}}n^{-m}r_n^{d(m-1)},
\end{align*}
for $n$ large enough so that ${\binom{n-m}{m-j}}/{\binom{n}{m}} \leq 2^j n^{-j}$ and $nr_n^d \leq 1$. We deduce that
\begin{align*}
\Var \left ( \frac{G_n(s,t)}{r_n^{d(m-1)}\binom{n}{m}} \right ) & = \frac{1}{r_n^{2d(m-1)}} \Var(U_n) \leq C_{d,m,t^+,\|f\|_{\infty}} (n^m r_n^{d(m-1)})^{-1} \\
& \leq C_{d,k,t^+,\|f\|_{\infty}} (n^{k+2} r_n^{d(k+1)})^{-1}.
\end{align*}
Bounding the variance of $L_{r_nt}$ proceeds from the same calculation, noting that $L_{r_nt}$ is a $U$-statistic of order $m=k+3$. Namely, proceeding as above leads to
\begin{align*}
\Var \left ( \frac{L_{r_nt}}{\binom{n}{m}} \right ) \leq C_{d,m,t^+,\|f\|_{\infty}}n^{-m}r_n^{d(m-1)},
\end{align*}
with $m=k+3$, so that
\begin{align*}
\Var \left ( \frac{L_{r_nt}}{(r_n^d)^{k+1}\binom{n}{k+2}} \right ) & \leq C_{d,k,t^+,\|f\|_{\infty}}\frac{\binom{n}{k+3}^2}{\binom{n}{k+2}^2 r_n^{d(2k+2)}}n^{-(k+3)}r_n^{d(k+2)} \\
& \leq C_{d,k,t^+,\|f\|_{\infty}} \frac{n r_n^d}{n^{k+2}r_n^{d(k+1)}} \leq C_{d,k,t^+,\|f\|_{\infty}}  (n^{k+2}r_n^{d(k+1)})^{-1},
\end{align*}
for $n$ large enough.

\subsection*{End of the proof}

Let $k \leq d-4$ and choose $r_n=n^{-\frac{k+2}{2+d(k+1)}}$. It holds 
\begin{align*}
r_n^2 & = n^{-(k+2)}r_n^{-d(k+1)} = n^{-\frac{2(k+2)}{2+d(k+1)}}, \\
n r_n^{d} & = n^{\frac{2 + d(k+1) - d(k+2)}{2 + d(k+1)}} = n^{\frac{2-d}{2 + d(k+1)}} \leq r_n.
\end{align*}
The above calculation then leads to, for any $0 < s\leq t \leq u \leq v \leq t^+$, and $n$ large enough, 
\begin{align*}
\E \left[(\xi_{k,n}-\mu_k)(R_{s,t,u,v})^2\right] & \leq \E \left[(\xi_{k,n}-\mu_k)(R_{s,t,u,v})\right]^2 + \Var \left[ (\xi_{k,n}-\mu_k)(R_{s,t,u,v}) \right] \\ 
& \leq C_{d,k,t^+, \|f\|_{\infty}}n^{-(k+2)}r_n^{-d(k+1)} + C_{d,k,t^+,\|f\|_{\infty}} (L \vee \frac{\|f\|_{\infty}}{\tau_{\min}})^2r_n^2 \\
& \leq C_{k,d,t^+, \|f\|_{\infty}, \tau_{\min}, L}n^{- \frac{2(k+2)}{2+d(k+1)}}.
\end{align*}
Now, for $k \geq d-4$ and $r_n=n^{-\frac{k+4}{d(k+3)}}$, we get
\begin{align*}
n^2r_n^{2d} & = n^{-(k+2)}r_n^{-d(k+1)} = n^{-\frac{2}{k+3}}, \\
r_n & = n^{-\frac{k+4}{d(k+3)}} \leq n^{-\frac{1}{k+3}} = nr_n^d. 
\end{align*}
This yields, for $n$ large enough, 
\begin{align*}
\E \left[(\xi_{k,n}-\mu_k)(R_{s,t,u,v})^2\right] \leq C_{k,d,t^+, \|f\|_{\infty}, \tau_{\min}, L}n^{- \frac{2}{k+3}}.
\end{align*}

\subsection{Proof of Corollary \ref{cor:concentration}}
\label{proof:cor_conc}

Assume without loss of generality that the points are sampled according to $f_1$. Let $0 \leq s \leq t \leq u \leq v \leq \infty$. For $i = 1, 2$, denote by 
\[
l_i = \frac{\int_{\mathcal{M}} f_i^{k+2}\mathrm{d}\mathcal{H}}{(k+2)!} \int_{(\mathbb{R}^d)^{k+1}} H_{s, t, u, v} (0, y_1, \ldots, y_{k+1}) dy_1 \ldots dy_{k+1}.
\]

By Chebyshev's inequality, 

\[ \mathbb{P} \left( |\xi_{k, n}(R_{s, t, u, v}) - l_1 | \geq \frac{|l_1 - l_2|}{2} \right) \leq \frac{4 \E \left[(\xi_{k,n}(R_{s,t,u,v})-l_1)^2\right]}{|l_1-l_2|^2}.
\]

Inverting the above formula and using the variance bound in the proof of Theorem \ref{theo:Owada} yields that there exists a constant $C$ such that with probability greater or equal than $1-C\frac{n^{-\frac{2(k+2)}{d(k+1)}}}{(\int_M |f_1^{k+2}-f_2^{k+2}|)^2}$,

\[ |\xi_{k,n} - l_1| \geq \frac{|l_1 -l_2|}{2}.
\]

This means that with at least the same probability, the data are correctly labeled as being sampled according to $f_1$.


\bibliographystyle{imsart-number} 
\bibliography{bibliographie.bib}       


\end{document}